\documentclass[11pt,a4paper]{article}
\usepackage[noblocks]{authblk}
\usepackage{times}
\usepackage{a4wide}
\title{Modular Bond-graph Modelling and Analysis\\ of
  Biomolecular Systems}
\author{Peter J. Gawthrop\footnote{Corresponding author. \textbf{peter.gawthrop@unimelb.edu.au}}}
\affil{
  Systems Biology Laboratory,
  Melbourne School of Engineering,
  University of Melbourne,
  Victoria 3010, Australia.
  \authorcr
  Department of Electrical and Electronic Engineering, 
  Melbourne School of Engineering,
  University of Melbourne,
  Victoria 3010, Australia.
  }

\author{Edmund J. Crampin}
\affil{
   Systems Biology Laboratory,
   Melbourne School of Engineering,
   University of Melbourne,
   Victoria 3010, Australia.
   \authorcr 
   School of Mathematics and Statistics,
   University of Melbourne,
   Victoria 3010, Australia.
   \authorcr 
   School of Medicine,
   University of Melbourne,
   Victoria 3010, Australia.
  \authorcr 
   ARC Centre of Excellence in Convergent Bio-Nano Science,
   Melbourne School of Engineering,
   University of Melbourne,
   Victoria 3010, Australia.
  \authorcr
  Centre for Systems Genomics,
  University of Melbourne,
  Victoria 3010, Australia.
}

\usepackage{amsmath,amssymb,amscd,amstext,mathtools,extarrows}

\newcommand{\lb}{\left (}
\newcommand{\pd}[2]{\frac{\partial #1}{\partial{#2}}}
\newcommand{\diag}{\text{diag }}

\newcommand{\Ln}{\text{\bf Ln }}
\newcommand{\Exp}{\text{\bf Exp }}
  
\newcommand{\rb}{\right )}
\newcommand{\ddt}[1]{\dot{#1}}
\newcommand{\dx}{\ddt{x}}
\newcommand{\jw}{j\omega}

\newcommand{\Nf}{{N^f}}
\newcommand{\Nr}{{N^r}}
\newcommand{\Nfr}{{N^{fr}}}
\newcommand{\Ncd}{{N^{cd}}}
\newcommand{\Gcd}{{G^{cd}}}
\newcommand{\Lcd}{{L^{cd}}}

\newcommand{\sss}[1]{\bar{#1}}
\newcommand{\ttt}[1]{\tilde{#1}}
\newcommand{\lin}[1]{\breve{#1}}

\newcommand{\vt}{\ttt{v}}
\newcommand{\kt}{\lin{\kappa}}
\newcommand{\Kt}{\lin{K}}

\newcommand{\xs}{\sss{x}}

\renewcommand{\AA}{\boldsymbol{A}}
\newcommand{\BB}{\boldsymbol{B}}
\newcommand{\CC}{\boldsymbol{C}}

\newcommand{\GG}{\boldsymbol{G}}
\newcommand{\AAA}{\boldsymbol{a}}
\newcommand{\BBB}{\boldsymbol{b}}
\newcommand{\CCC}{\boldsymbol{c}}
\newcommand{\DDD}{\boldsymbol{d}}

\newcommand{\KK}{\boldsymbol{K}}

\newcommand{\VV}{\boldsymbol{V}}
\newcommand{\UU}{\boldsymbol{U}}
\newcommand{\XX}{\boldsymbol{X}}
\newcommand{\xx}{\boldsymbol{x}}

\newcommand{\II}{\mathcal{I}}

\newcommand{\kkappa}{\boldsymbol{\kappa}}
\newcommand{\mmu}{\boldsymbol{\mu}}
\newcommand{\nmu}{{\check{\mu}}}
\newcommand{\nmmu}{{\boldsymbol{\check{\mu}}}}

\newcommand{\nA}{{\check{A}}}
\newcommand{\nAA}{{\boldsymbol{\check{A}}}}
\newcommand{\mf}{\chi} 

\newcommand{\reacul}[2]{
  {\; \xrightleftharpoons[#2]{#1} \;}
}

\newcommand{\reacu}[1]{
  \reacul{#1}{}
}

\newcommand{\reac}{
  \reacu{}
}

\usepackage{siunitx}
\newcommand{\Si}[1]{~(\si{#1})}

\usepackage{BG}

\usepackage[numbers,compress,sort]{natbib}
\usepackage{times}

\usepackage{listings}

\usepackage{graphicx,subfigure,fancybox,color}

\newcommand{\SubFig}[3]{
  \subfigure[#2]{
    \includegraphics[width=#3\linewidth]{Figs/#1.pdf}
    \label{subfig:#1}
  }
}

\usepackage[pagebackref]{hyperref}
\hypersetup{colorlinks=true,citecolor=blue,linkcolor=blue,urlcolor=green}
\usepackage{doi}
\usepackage{url}

\begin{document}
\maketitle
\begin{abstract}
  Bond graphs can be used to build thermodynamically-compliant
  hierarchical models of biomolecular systems. As bond graphs have been
  widely used to model, analyse and synthesise engineering systems, this
  paper suggests that they can play the same r\^{o}le in the
  modelling, analysis and synthesis of biomolecular systems. 
  The particular structure of bond graphs arising from biomolecular
  systems is established and used to elucidate the relation between
  thermodynamically closed and open systems.
  Block diagram representations of the dynamics implied by these bond
  graphs are used to reveal implicit feedback structures and are
  linearised to allow the application of control-theoretical methods.
  
  Two concepts of modularity are examined: computational modularity
  where physical correctness is retained and behavioural
  modularity where module behaviour (such as ultrasensitivity) is
  retained.
  As well as providing computational modularity, bond graphs provide a
  natural formulation of behavioural modularity and reveal the sources
  of retroactivity. A bond graph approach to reducing retroactivity,
  and thus inter-module interaction, is shown to require
  a power supply such as that provided by the $ATP \reac ADP + Pi$
  reaction.
  
  The MAPK cascade (Raf-MEK-ERK pathway) is used as an illustrative
  example.
\end{abstract}
\newpage

\section{Introduction}
\label{sec:introduction}
In their review paper \emph{The r\^{o}le of control and system theory
  in systems biology}, \citet{WelBulKalMasVer08} suggest that
``systems biology is an area where systematic methods for model
development and analysis, such as bond graphs, could make useful new
contributions as they have done in the physical world''. 
The purpose of this paper is to show that bond graphs not only provide
a systematic methods for model development and analysis of
biomolecular systems but also provide a bridge allowing application of
control engineering methodology, in particular feedback concepts, to
systems biology.

Bond graphs were introduced by \citet{Pay61} and their engineering
application is described in  number of
text books \citep{Wel79,GawSmi96,MukKarSam06,KarMarRos12} and a
tutorial for control engineers \citep{GawBev07}.
%
Bond graphs were first used to model chemical reaction networks by
\citet{OstPerKat71} and a detailed account is given by
\citet{OstPerKat73}.
Subsequent to this, the bond graph approach to chemical reactions has
been extended by \citet{Cel91}, \citet{ThoMoc06} and \citet{GreCel12}.
%
More recently, the bond graph approach has been used to analyse
biochemical cycles by \citet{GawCra14} and has been shown to provide
a modular approach to building hierarchical biomolecular system models
which are robustly thermodynamically compliant \citep{GawCurCra15}; combining
thermodynamically compliant modules gives a thermodynamically
compliant system.
In this paper we will call this concept \emph{computational
  modularity}.

Computational modularity is a necessary condition for building
physically correct computational models of biomolecular systems.
However, computational modularity does not imply that module
properties (such as ultrasensitivity) are retained when a module is
incorporated into a larger system. In the context of
engineering, modules often have buffer amplifiers at the interface so
that they have unidirectional connections and may thus be represented
and analysed on a block diagram or signal flow graph where the
properties of each module are retained.  This will be called
\emph{behavioural modularity} in this paper.
However, biological networks do not usually have this unidirectional
property but rather display \emph{retroactivity}
\citep{SaeKreCon04,SaeKreGil05,VecNinSon08,OssVenMer11,Vec13,VecMur14};
retroactivity modifies the properties of the interacting modules.  As
will be shown, the property of retroactivity is naturally captured by
bond graphs.
In particular, a bond graph approach to reducing retroactivity, and
thus inter-module interaction, is discussed and shown to require a
power supply such as that provided by the $ATP \reac ADP + Pi$
reaction.
%

Early attempts at modelling the MAPK cascade \citep{HuaFer96,Kho00},
used modules which displayed behavioural modularity. However,
because they use the Michaelis-Menten
approximation, the modules do not have the property of computational
modularity and thus the results were based on a non-physical
model. This was noted in later work which examined the neglected interactions: in particular,
\citet{OrtAceWes02} show that ``product dependence and bifunctionality
compromise the ultrasensitivity of signal transduction cascades''
and the ``effects of sequestration on signal transduction cascades''
are considered by \citet{BluBruLeg06}. In this paper, the MAPK
cascade is used as an illustrative example which illustrates how a
computationally modular approach based on bond graphs avoids the
errors associated with assuming irreversible Michaelis-Menten
kinetics. Moreover, the bond graph approach to reducing retroactivity
is used to make the modules approximately modular in the behavioural
sense. This emphasises the necessity for a power supply to support
signalling networks in biology as well as in engineering.

The bond graph approach gives the set of \emph{nonlinear} ordinary
differential equations describing the biomolecular system being modelled.
Linearisation of non-linear systems is a standard technique in control
engineering: as discussed by \citet{GooGraSal01}, ``The incentive to
try to approximate a nonlinear system by a linear model is that the
science and art of linear control is vastly more complete and simpler
than they are for the nonlinear case.''. Nevertheless, it is important
to realise that conclusions drawn from linearisation can only be
verified using the full \emph{nonlinear} equations.
In the context of bond graphs, linearisation (and the associated
concept of sensitivity) has been treated by a number of authors
\citep{Kar77,Gaw00c,Bor11a}. This paper builds on this work to
explicitly derive the bond graph  corresponding to the
linearised nonlinear system and thus provide a method to analyse
behavioural modularity.

\S~\ref{sec:bond-graph-modelling} briefly shows how biomolecular
systems can be modelled using bond graphs. \S~\ref{sec:closed-systems}
shows how thermodynamically closed systems can be converted to
thermodynamically open systems using the twin notions of chemostats
and flowstats. Linearisation is required to understand module
behaviour, and this is developed in
\S~\ref{sec:linearisation}. \S~\ref{sec:feedback} looks at modularity,
retroactivity and feedback and \S~\ref{sec:map-kinase-cascades}
illustrates the main results using the MAPK cascade
example. \S~\ref{sec:conclusion} concludes the paper and suggests
future research directions.

\section{Bond Graph  Modelling of Biomolecular Systems}
\label{sec:bond-graph-modelling}
\begin{figure}[htbp]
  \centering
  \SubFig{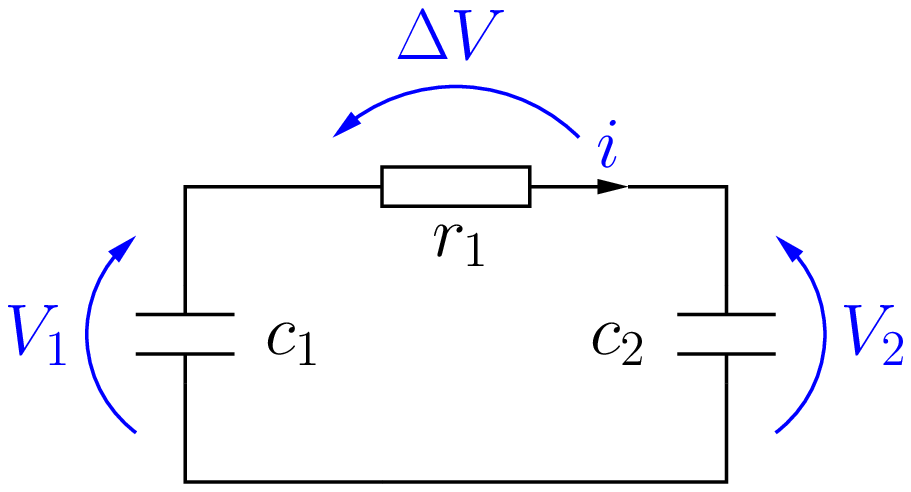}{Electrical Circuit}{0.3}
  \SubFig{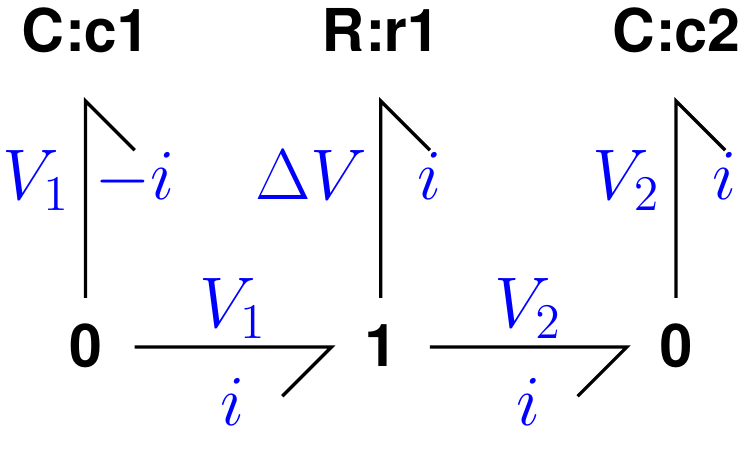}{Bond graph}{0.3}
  \SubFig{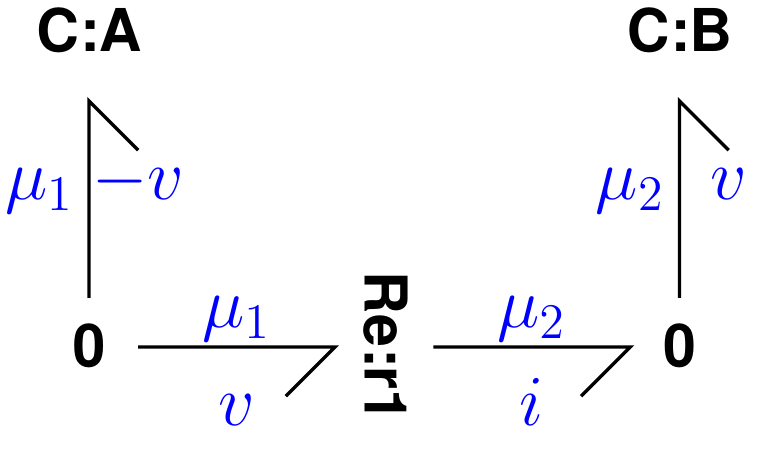}{Bond graph for $A \reacu{1} B$}{0.3}
  \caption{Simple example. (a) An RC circuit schematic. (b) As
    discussed in the text, capacitors ($c_1$ \& $c_2$) are represented by bond graph \C
    components, resistors ($r_1$) by bond graph \R components and connections
    by \zero and \one junctions; the bonds ($\rightharpoondown$) carry
    the effort (voltage) and flow (current) variables. (c) The
    reaction $A \reacu{1} B$. As discussed in
    \S~\ref{sec:bond-graph-comp}, the \one and \R components are
    replaced by a single two port \Re component representing the
    reaction and the \C components correspond to the reactants $A$ and
    $B$.}
  \label{fig:CRC}
\end{figure}

As discussed by \citet{Max71}, the use of ``mathematical or formal
analogy'' enables us to avail ``ourselves of the mathematical labours
of those who had already solved problems essentially the same.'' The
bond graph approach provides a systematic approach to the use of
analogy in the modelling of systems across different physical domains;
in the context of this paper, this allows engineering concepts to be
carried across to biomolecular systems. 

A number of text books about bond graphs
\citep{Wel79,GawSmi96,MukKarSam06,KarMarRos12} and a tutorial for
control engineers \citep{GawBev07} are available.
Briefly, bond graphs focus on a pair of variables generically termed
\emph{effort} $e$ and \emph{flow} $f$ whose product is power
$p=ef$. In the electrical domain, effort is identified with voltage
$V$\Si{V} and flow with current $i$\Si{C.sec^{-1}} and in the
mechanical domain effort is identified with force $F$\Si{N} and flow
with velocity $v$\Si{m.sec^{-1}}. Thus voltage and force are effort
analogies and current and velocity are flow analogies. Although the
effort (and the flow) variables have different units in each domain,
their product (power) has the same units ($\si{W}$ or
\si{J.sec^{-1}}); power is the common currency of disparate physical
domains. The pair $e$ $f$ is represented on the bond graph by the
harpoon symbol:
$\rightharpoondown$
which can be optionally annotated with specific effort and flow
variables, for example
$\xrightharpoondown[f]{e}$.
Sign convention is handled by the harpoon direction: thus if $e$ and
$f$ are positive, the flow $f$ is in the harpoon direction.

As well as analogous variables, bond graphs deal in analogous
\emph{components}. Thus the bond graph \C component models both the
ideal electrical capacitor (with capacitance $c_c$) and the ideal
mechanical spring (with stiffness $K_s$). In both
cases, the \C component physically accumulates flow to give the
integrated flow $q$ corresponding to electrical charge or mechanical
displacement. In the linear case, this gives an \emph{effort}
proportional to $q$. To summarise:
\begin{xalignat}{3}
  \dot{Q} &= i & V &= \frac{Q}{c_c} & &(electrical)\notag\\
  \dot{x} &= v & F &= K_s x & &(mechanical)\notag\\
  \dot{q} &= f & e &= K q & &(generic) \label{eq:C}
\end{xalignat}

Similarly, electrical resistors and mechanical dampers are represented
by bond graph  \R components where:
\begin{xalignat}{2}
  V &= r i & &(electrical)\\
  F &= r v & &(mechanical)\\
  e &= r f & &(generic)
\end{xalignat}
where $r$ represents the (linear) electrical resistance  and
mechanical damping factor.

Bonds are connected by \zero and \one junctions which again conserve
energy; the \zero junction gives the same effort on each impinging
bond and the \one junction gives the same flow on each impinging
bond.

Figure \ref{fig:CRC} shows a simple electrical circuit connecting two
capacitors with capacitance $c_1$ and $c_2$ by a resistor with
resistance $r_1$. Figure \ref{subfig:CRC_pic} gives the electrical
schematic diagram and Figure \ref{subfig:CRC_abg} gives the
corresponding bond graph  which uses the \C, \R, \zero and \one
components connected by bonds.

The bond graph  \TF component represents both an electrical
transformer and a mechanical lever with ratio $\rho$. In generic terms,
the bond graph  fragment:
$\xrightharpoondown[f_1]{e_1}$ \BTF{\rho} $\xrightharpoondown[f_2]{e_2}$
represents the two equations:
\begin{xalignat}{2}
  f_2 &= \rho f_1 & e_1 &= \rho e_2
\end{xalignat}
Note that energy is conserved as
\begin{equation}
  p_2 = e_2  f_2 = \frac{e_1}{\rho} \rho f_1 = e_1 f_1 = p_1
\end{equation}

\subsection{Biomolecular bond graph  components}
\label{sec:bond-graph-comp}
It is assumed that biochemical reactions occur under conditions of
constant pressure (isobaric) and constant temperature
(isothermal). Under these conditions, the chemical potential $\mu_A$
of substance $A$ is given \cite{AtkPau11} in terms of its mole
fraction $\mf_A$ as:
\begin{equation} \label{eq:mu_A_0}
  \mu_A =  \mu_A^\star + RT \ln \mf_A \Si{~J.mol^{-1}}
\end{equation}
where the standard chemical potential $\mu_A^\star$ is the value of $\mu_A$ when $A$ is pure
($\mf_A=1$), $R = 8.314\Si{~JK^{-1}mol^{-1}}$ is the universal gas
constant, $T\Si{K}$ is the absolute temperature and $\ln$ is the
natural (or Napierian) logarithm\footnote{
  Unlike voltage and force (which could be dimensioned as
  \si{J.C^{-1}} and \si{J.m^{-1}} respectively) chemical potential
  does not have its own unit. \citet{JobHer06} suggest Gibbs (\Si{G})
  as the the unit of chemical potential.
}.
It is convenient to define a \emph{normalised} chemical potential
$\nmu_A$ as:
  \begin{equation}
    \label{eq:nmu}
    \nmu_A = \frac{\mu_A}{RT}
  \end{equation}
%

The key to modelling chemical reactions by bond graphs is to determine
the appropriate effort and flow variables.  As discussed by
\citet{OstPerKat71,OstPerKat73}, the appropriate \emph{effort}
variable is \emph{chemical potential} $\mu$ and the appropriate
\emph{flow} variable is molar flow rate $v$. 

In the context of chemical reactions, the bond graph \C component of
Equation \eqref{eq:C} is defined by Equation \eqref{eq:mu_A_0} as:
\begin{xalignat}{3}\label{eq:C_chem}
  \dot{x}_A &= v_A \Si{~mol.sec^{-1}} & \mu_A &= RT \ln K_A x_A \Si{~J.mol^{-1}}& &(chemical)
\end{xalignat}
where $x_A$ is the molar amount of $A$ and the \emph{thermodynamic
  constant} $K_A$ is given by
\begin{equation}
  \label{eq:K_A}
  K_A = \frac{1}{n_{total}}\exp{\frac{\mu_A^\star}{RT}} 
  = \frac{1}{n_{total}}\exp{\nmu_A^\star} \Si{~mol^{-1}}
\end{equation}
where $n_{total}$ is the total number of moles in the mixture.
Alternatively, \eqref{eq:C_chem} can be written more simply in terms of the
normalised chemical potential $\nmu$ of Equation \eqref{eq:nmu}:
\begin{xalignat}{2}\label{eq:C_chem_norm}
  \dot{x}_A &= v_A \Si{~mol.sec^{-1}} & \nmu_A &=  \ln K_A x_A
\end{xalignat}
We follow \citet{OstPerKat73} in describing chemical reactions in
terms of the \emph{Marcelin -- de Donder} formulae as discussed by
\citet{Rys58} and \citet{GawCra14}. In particular, given the $i$th reaction
\citep[(5.9)]{OstPerKat73}:
\begin{equation}
  \label{eq:react}
  \nu^f_A A + \nu^f_B B + \nu^f_C C + \nu^f_D D \dots 
  \reacu{i} 
  \nu^r_A A + \nu^r_B B + \nu^r_C C + \nu^r_D D \dots 
\end{equation}
where the stoichiometric coefficients $\nu$ are either zero or
positive integers, the \emph{forward affinity} $A^f_i$ and the
\emph{reverse affinity} $A^r_i$ are defined as:
\begin{align}
  A^f_i &= \nu^f_A \mu_A + \nu^f_B \mu_B + \nu^f_C \mu_C + \nu^f_D \mu_D \dots  \label{eq:A^f}\\
  A^r_i &= \nu^r_A \mu_A + \nu^r_B \mu_B + \nu^r_C \mu_C + \nu^r_D \mu_D \dots  \label{eq:A^r}
\end{align}
The units of affinity are the same as those of chemical potential: $\si{J.mol^{-1}}$.
Again, normalised affinities are useful:
\begin{xalignat}{2}
  \label{eq:A_norm}
  \nA^f_i &= \frac{A^f_i}{RT} &
  \nA^r_i &= \frac{A^r_i}{RT} 
\end{xalignat}

The $i$th reaction flow $v_i$ is then given by:
\begin{xalignat}{3}
 v_i &=  \kappa_i \lb v^+_0 - v^-_0 \rb&
   \text{where }
   v^+_0 &=   e^\frac{A^f_i}{RT} = e^{\nA^f_i}&
   \text{and }
   v^-_0 &=  e^\frac{A^r_i}{RT}\ = e^{\nA^r_i} \label{eq:v_exp}
\end{xalignat}
Note that the
arguments of the exponential terms are dimensionless as are $v^+_0$
and $v^-_0$. The units of the reaction rate constant $\kappa_i$ are
those of molar flow rate: $\si{mol.sec^{-1}}$.

The $i$th reaction flow $v_i$ depends on the forward and reverse affinities
$A^f_i$ and $A^r_i$ but cannot be written as the difference between the
affinities. Unlike the electrical $\R$ component (see Figure
\ref{fig:CRC}), it cannot be written as a one port component with the
flow dependent on the difference between the efforts. However, as
discussed by \citet{GawCra14}, a two port resistive component, the \Re
component, can be used to model the reaction \eqref{eq:v_exp}.

The fact that the capacitive \C and resistive \Re components are
intrinsically non-linear is one factor distinguishing biochemical
systems from the electrical and mechanical systems of Equation \eqref{eq:C}.

The \TF component is used in this context to account for any non-unity and
non-zero stoichiometric coefficients $\nu$ in Equation
\eqref{eq:react} \citep{OstPerKat71,OstPerKat73,GawCra14}.
Moreover, as will be discussed in the next section, the \TF component can be
used to abstract the entire network of bonds, \zero and \one junctions
connecting the \C and \Re components.

\subsection{Examples}
\label{sec:model_examples}
Consider the simple reaction $A \reacu{1} B$. In this case $\nA^f =
\nmu_A$ and $\nA^r = \nmu_B$. With reference to Figure
\ref{subfig:CRCRe_abg}, substance $A$ is modelled by \BC{A}, substance
$B$ is modelled by \BC{B} and the reaction by \BRe{r1}.  The equations
of the \C components correspond to Equation \eqref{eq:C_chem} and
that of the \Re component to \eqref{eq:v_exp}.
The equations are:
\begin{xalignat}{2}
  \nA^f = \nmu_A &= \ln K_A x_A &
  \nA^r = \nmu_B &= \ln K_B x_B \\
  v^+_0 &= e^{\nA^f} = K_A x_A &
  v^-_0 &= e^{\nA^r} = K_B x_B \label{eq:simple_v_0}\\
  v &= \kappa \lb K_A x_A -  K_B x_A \rb = k^+ x_A-k^-x_B\label{eq:simple_v}&
\text{where } k^+ &= \kappa K_A \text{ and }  k^- = \kappa K_B 
\end{xalignat}
In this simple case the equations are linear and the rate constants
$k^+$ and $k^-$ are given in terms of the reaction rate-constant $\kappa$ and the thermodynamic
constants $K_A$ and $K_B$. The equilibrium constant $K^{eq}$ is given by:
\begin{equation}
  \label{eq:Keq}
  K^{eq} = \frac{k^+}{k^-} = \frac{K_A}{K_B}
\end{equation}
and is thus a function of the thermodynamic constants $K_A$ and $K_B$
but not the reaction rate-constant $\kappa$. A general formula
relating all the equilibrium constants in a biomolecular network to
the rate constants is given by \citet[\S~3]{GawCurCra15}.

\begin{figure}[htbp]
  \centering
  \SubFig{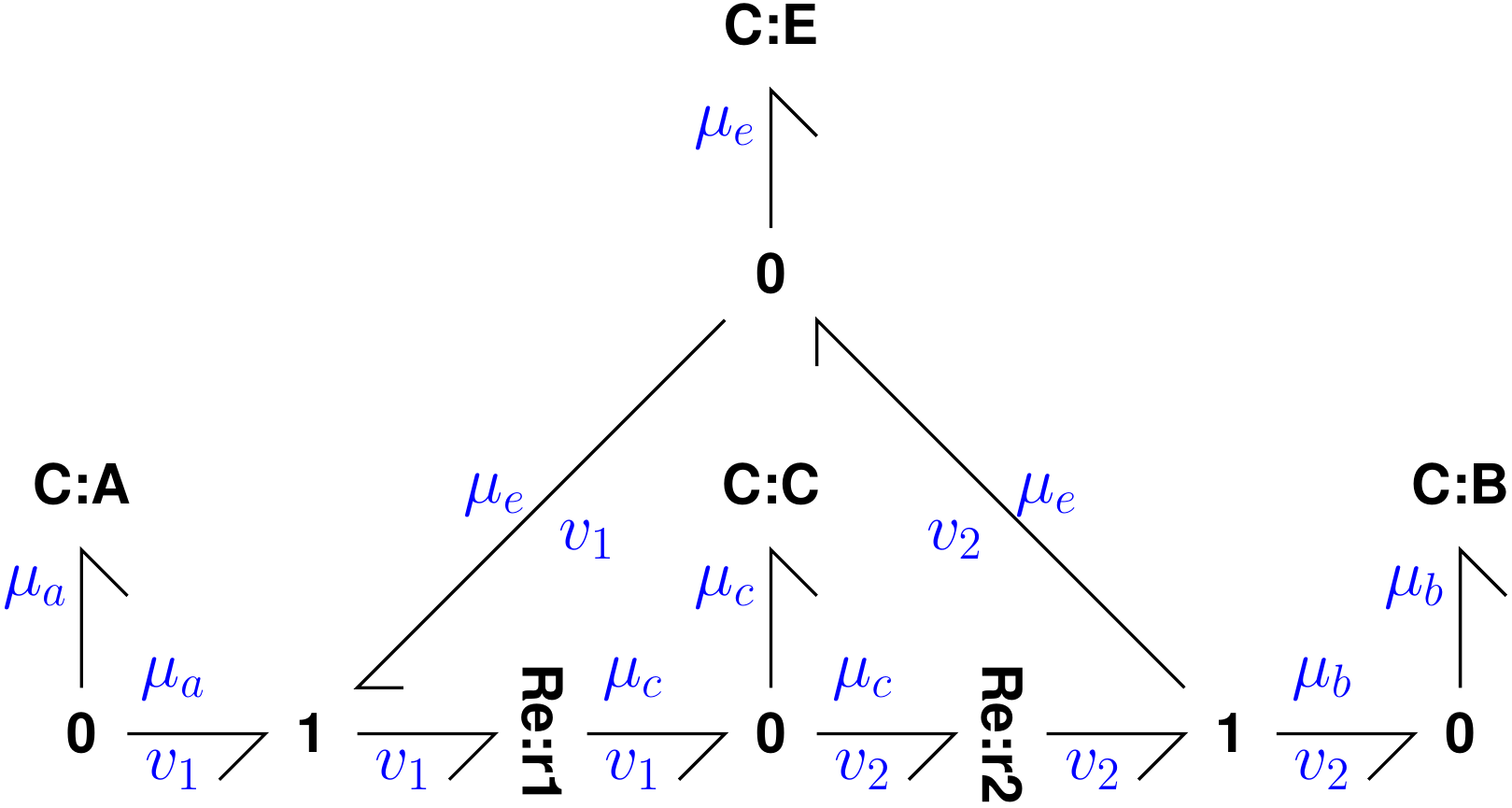}{Bond graph}{0.45}
  \SubFig{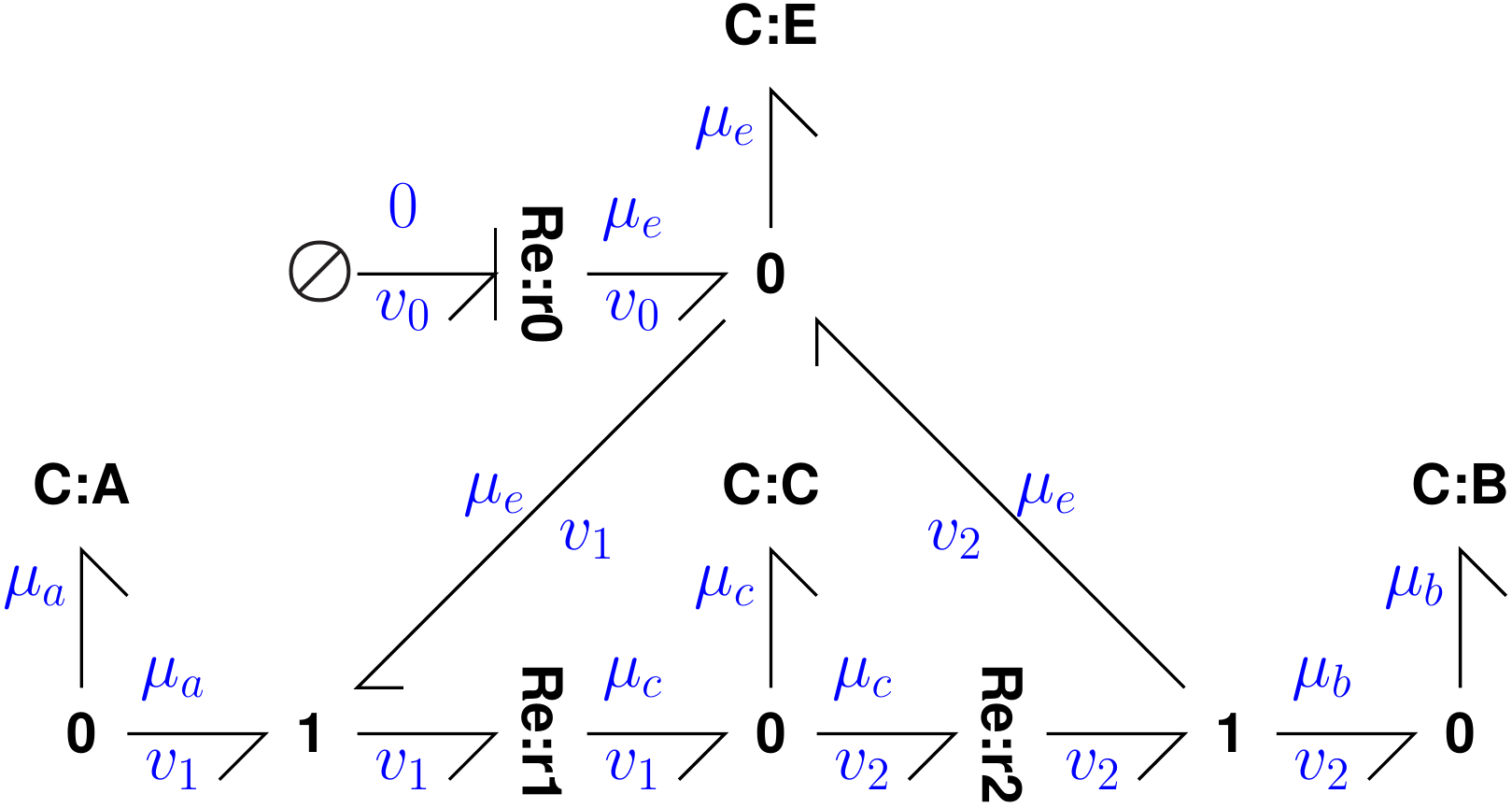}{Bond graph with enzyme degradation}{0.45}
  \caption[Example: Enzyme catalysed reaction]
  {Example: Enzyme catalysed reaction.
    (a) Bond graph of the enzyme catalysed reaction $A+E \reacu{1} C
    \reacu{2} B+E$. $A$ is the reactant, $B$ the product, $C$ the
    intermediate complex and $E$ the enzyme. ($A$ and $B$ are used as
    chemostats in \S~\ref{sec:closed-systems}).
    (b) An enzyme degradation reaction \BRe{r0} is added. (\BRe{r0} is used
    as a flowstat in \S~\ref{sec:closed-systems}).
}
\label{fig:ECR}
\end{figure}
The enzyme catalysed reaction
\begin{equation}
  \label{eq:ECR}
  A+E \reacu{1} C \reacu{2} B+E
\end{equation}
where $A$ is the reactant, $B$ the product, $C$ the intermediate
complex and $E$ the enzyme, is ubiquitous in biochemical systems. The
reaction \eqref{eq:ECR} was first modelled using bond graphs by
\citet[Fig. 5.9]{OstPerKat73}. 

Figure \ref{subfig:ABEC0_abg} is the bond graph corresponding to
Equation \eqref{eq:ECR}. The components \BRe{r1} and \BRe{r2}
represent reactions $\reacu{1}$ and $\reacu{2}$ and the four \C
components \BC{A}, \BC{B}, \BC{C} and \BC{E} represent the four
species $A$, $B$, $C$ and $E$. The left-hand \one junction ensures that the
flow out of \BC{A} and \BC{E} is the reaction flow $v_1$ 
and the right-hand \one junction ensures that the flow into  \BC{B}
and \BC{E} is the reaction flow $v_2$. The net flow into \BC{E} is
thus $v_1-v_2$.

The additional reaction \BRe{r0} has been added in Figure
\ref{subfig:ABEC_abg} together with the \emph{zero-potential} source
$\boldsymbol{\oslash}$; this can be used to model enzyme degradation.
\BC{A} and \BC{B} are used in \S~\ref{sec:closed-systems} as an
example of a \emph{chemostat} and \BRe{r0} as an example of a
\emph{flowstat}. 
The enzyme catalysed reaction is analysed further in
\S~\ref{sec:example-modul-enzyme}.

\section{Closed systems and open systems: chemostats and flowstats}
\label{sec:closed-systems}
\begin{figure}[htbp]
  \centering
  \SubFig{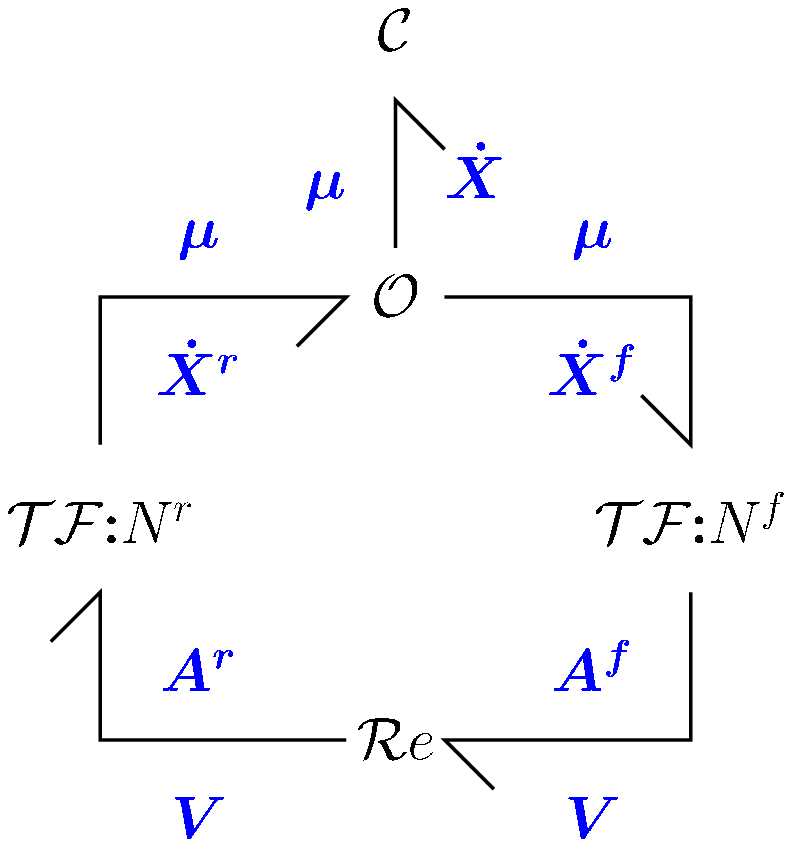}{Closed system bond graph}{0.3}
  \SubFig{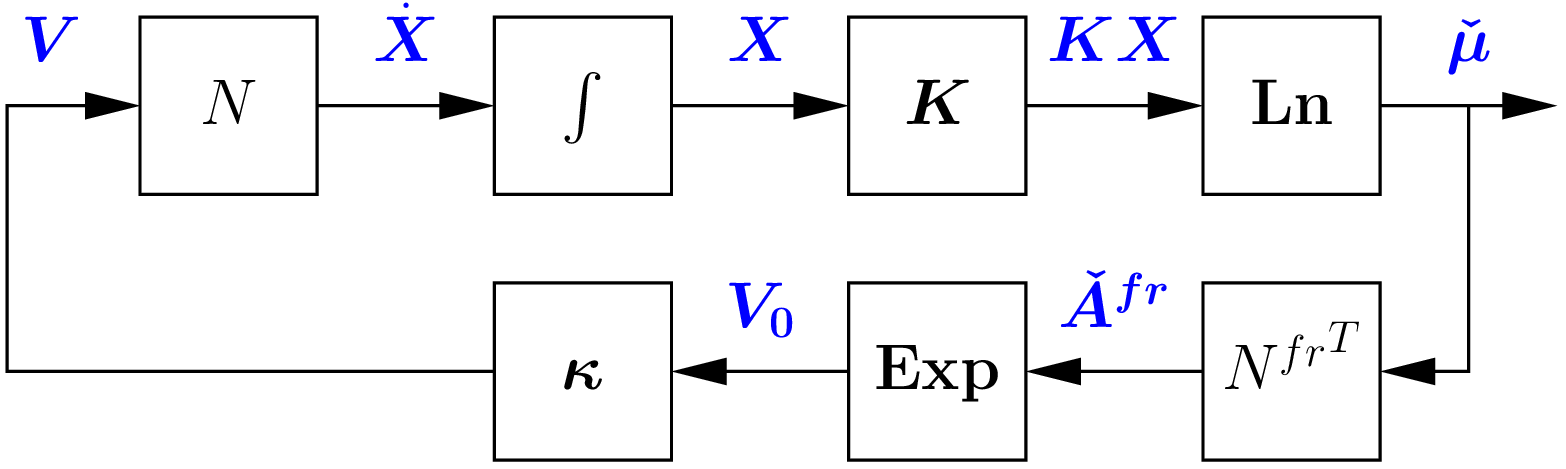}{Closed system block-diagram}{0.6}
  \SubFig{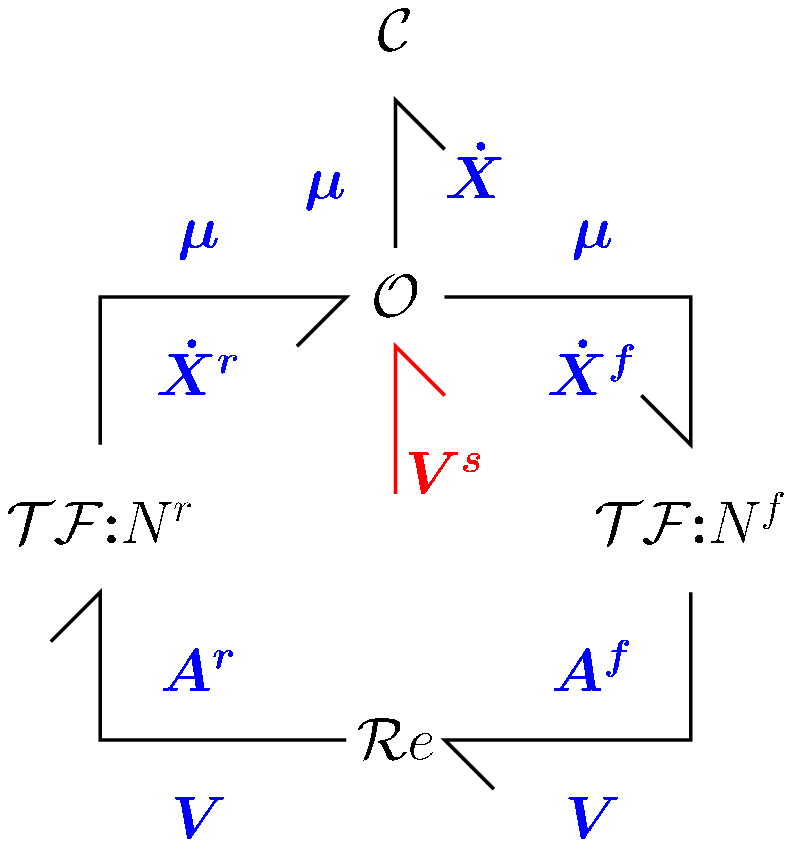}{Open system bond graph}{0.3}
  \SubFig{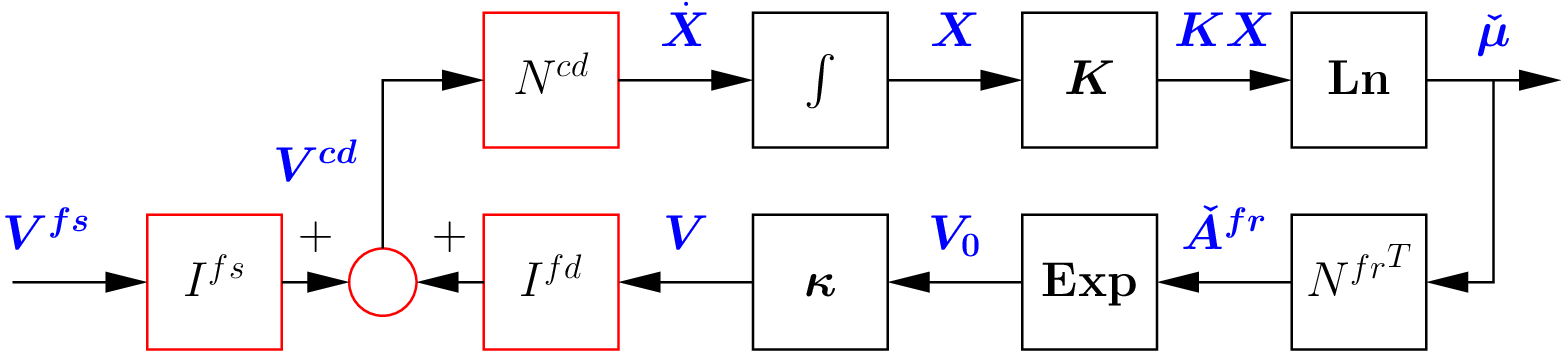}{Open system block diagram}{0.65}
  \caption[Closed \& open systems]{
    Closed \& open systems (a) General closed system. The bond symbols
    $\rightharpoondown$ correspond to \emph{vectors} of bonds;
    $\mathcal{C}$, $\mathcal{R}e$ and $\mathcal{O}$ correspond to
    arrays of \C, \Re and \zero components ; the two $\mathcal{TF}$
    components represent the intervening junction structure comprising
    bonds, \zero and \one junctions and \TF components. $N^f$ and
    $N^r$ are the forward and reverse stoichiometric matrices. (b) The
    corresponding block diagram.
    (d) The addition of the chemostat \& flowstat flows $V^{s}$ to the
    \emph{closed} system of Figure \ref{subfig:Closed_bg} gives an \emph{open}
    system. 
    (d) The corresponding block diagram.
}
   \label{fig:Closed}
\end{figure}

Specific bond graphs (such as Figures \ref{subfig:CRCRe_abg} and
\ref{subfig:ABEC0_abg}) model specific sets of chemical reactions. It
is convenient to generalise such bond graphs to allow generic
statements to be made and generic equations to be written.  The molar
amounts of the $n_X$ species $x_A,x_B, \dots$, the corresponding
chemical potentials $\mu_A,\mu_B, \dots$  and the corresponding
thermodynamic constants $K_A,K_B, \dots$ are collected
into column vectors:
\begin{xalignat}{3}
  \label{eq:XX}
  \XX &=
  \begin{pmatrix}
    x_A\\x_B\\\vdots
  \end{pmatrix}&
  \mmu &=
  \begin{pmatrix}
    \mu_A\\\mu_B\\\vdots
  \end{pmatrix}&
  K &=
  \begin{pmatrix}
    K_A\\K_B\\\vdots
  \end{pmatrix}
\end{xalignat}
Similarly,the $n_V$ reaction flows $v_1,v_2,\dots$, affinities
(forward and reverse) $A_1,A_2,\dots$ and the
corresponding reaction constants $\kappa_1, \kappa_2, \dots$ are
collected into column vectors:
\begin{xalignat}{3}
  \label{eq:VV}
  \VV &=
  \begin{pmatrix}
    v_1\\v_2\\\vdots
  \end{pmatrix}&
  \AA &=
  \begin{pmatrix}
    A_1\\A_2\\\vdots
  \end{pmatrix}&
  \kappa &=
  \begin{pmatrix}
    \kappa_1\\\kappa_2\\\vdots
  \end{pmatrix}
\end{xalignat}
As discussed by \citet{KarMarRos12}, the \C components can be subsumed
into a single \emph{\C-field}, the \Re components (as two-port \R
components) subsumed into an \emph{R-field} and the connecting bonds,
\zero and \one junctions subsumed into a \emph{junction
  structure}. Moreover, as this junction structure transmits, but does
not store or dissipate energy, it can be modelled as the two multiport
transformers $\mathcal{TF}{:}N^f$ and $\mathcal{TF}{:}N^r$ shown in
Figure \ref{subfig:Closed_bg}.
These two multiport transformers are defined to transform \emph{flows}
as:
\begin{xalignat}{2}
  \dot{\XX}^r &= N^r \VV & \dot{\XX}^f &= N^f \VV
\end{xalignat}
Because they do not store or dissipate energy, it follows that the
\emph{affinities} are given by:
\begin{xalignat}{2}\label{eq:A}
  \AA^r &= \Nr^T \mmu & \AA^f &= \Nf^T\mmu
\end{xalignat}
As discussed by \citet{GawCra14}, and with reference to Figure
\ref{subfig:Closed_bg}, the system \emph{states} $\XX$ correspond to
the molar amounts of each species stored in each \C components and are given
in terms of the reaction flows $\VV$ as
\begin{xalignat}{2}
  \label{eq:Xdot}
  \dot{\XX} &= \dot{\XX}^r-\dot{\XX}^f = N \VV &
\text{where } N &= N^r-N^f
\end{xalignat}
$N$ is the \emph{stoichiometric matrix} \citep{Pal06}; $N^f$ and $N^r$ are referred
to as the forward and reverse stoichiometric matrices.

From Equation \eqref{eq:C_chem}, the composite chemical potential
$\mmu$
is given by the non-linear equation\footnote{
Following \citet{SchRaoJay13}, we use the convenient notation $\Exp \XX$
to denote the vector whose $i$th element is the exponential of the
$i$th element of $\XX$ and $\Ln \XX$ to denote the vector whose $i$th
element is the natural logarithm of the $i$th element of
$\XX$.}:
\begin{xalignat}{2} \label{eq:mmu}
  \frac{\mmu}{RT} &= \nmmu = \Ln \KK\XX &
\text{where } \KK &= \diag{K}
\end{xalignat}
and from Equation \eqref{eq:v_exp}, the composite reaction flow $\VV$ is
given by the non-linear equations
\begin{xalignat}{2}
  \VV^+_0 &= \Exp \lb \frac{\AA^{f}}{RT} \rb = \Exp \nAA^{f}&
  \VV^-_0 &= \Exp \lb \frac{\AA^{r}}{RT} \rb = \Exp \nAA^{r}\label{eq:V_0}\\
  \VV &= \kkappa \lb\VV^+_0 - \VV^-_0\rb &
  \text{where } \kkappa &= \diag{\kappa} \label{eq:V}
\end{xalignat}

Defining the composite stoichiometric and composite reaction constant
matrices $ N^{fr}$ and $\kkappa^{fr}$ as.
\begin{xalignat}{2}
  \label{eq:Nfr}
    N^{fr} &=
    \begin{pmatrix}
    N^f &  N^r
  \end{pmatrix} &
  \text{and } \kkappa^{fr} &=
  \begin{pmatrix}
    \kkappa &  -\kkappa
  \end{pmatrix} 
\end{xalignat}
Equations \eqref{eq:A}, \eqref{eq:V_0} and \eqref{eq:V} can be rewritten in a more
compact form as:
\begin{xalignat}{3}\label{eq:V_0_compact}
  \nAA^{fr} &= \Nfr^T \nmmu&
  \VV_0 &=  \Exp \nAA^{fr}&
  \VV &= \kkappa^{fr}\VV_0
\end{xalignat}
which can be combined to give a compact expression for the flows $\VV$
in terms of the state $\XX$
\begin{align}
  \VV  &= \kkappa^{fr} \Exp \lb \Nfr^T \Ln \KK \XX \rb   \label{eq:VV_compact}
\end{align}

\subsection{Block diagrams}
\label{sec:block-diagrams}
Block diagrams are the conventional way of describing systems in the
context of control design\citep{GooGraSal01}.
However, as discussed by \citet{GawBev07}, bond graphs are superior to block
diagrams in the context of system \emph{modelling}. Nevertheless, block
diagrams have advantages when \emph{analysing} the system dynamics
arising from the bond graph  model; in particular, block
diagrams expose the underlying feedback structure of the equations
arising from the bond graph model.
Figure \ref{subfig:Closed_bd_simp} is the block diagram corresponding to
the closed system bond graph of Figure \ref{subfig:Closed_bg}; it is a
diagrammatic way of writing down Equations \eqref{eq:Xdot}, \eqref{eq:mmu}
and \eqref{eq:V_0_compact}.  Each arrow corresponds to a vector
of signals corresponding to: the $n_X$ species concentrations $\XX$ and
normalised chemical potentials $\nmmu$, the $n_V$ reaction flows $\VV$
and the $2n_V$ normalised
forward and reverse affinities $\nAA^{fr}$. $\int$ represents the
integration of $\dot{X}$ to give $X$ implied by Equation
\eqref{eq:Xdot}.  $\Ln$ and $\Exp$ represent the nonlinear functions in
equations \eqref{eq:mmu} and \eqref{eq:V}.

\subsection{Examples}
\label{sec:ex_gen}
For example, in the case of the simple reaction $A \reacu{1} B$ of Figure \ref{subfig:CRCRe_abg}:
\begin{xalignat}{6}
  \label{eq:AB}
  \XX &=
  \begin{pmatrix}
    x_A\\x_B
  \end{pmatrix}&
  \VV &=
  \begin{pmatrix}
    v_1
  \end{pmatrix}&
  \Nf &= 
  \begin{pmatrix}
    1\\0
  \end{pmatrix}&
  \Nr &= 
  \begin{pmatrix}
    0\\1
  \end{pmatrix}&
  \Nfr &= 
  \begin{pmatrix}
    1&0\\0&1
  \end{pmatrix}&
  N &= 
  \begin{pmatrix}
    -1\\1
  \end{pmatrix} 
\end{xalignat}
As $\Nfr$ is a unit matrix, the ODE is
\begin{equation}
  -\dot{x}_A = \dot{x}_B = v_1 = \kappa \lb K_Ax_A -K_B x_B \rb
\end{equation}

In the case of the enzyme-catalysed reaction $A+E \reacu{1} C
\reacu{2} B+E$ of Figure \ref{subfig:ABEC_abg}:
\begin{xalignat}{5}
  \label{eq:ABEC}
  \XX &=
  \begin{pmatrix}
    x_A\\x_B\\x_C\\x_E
  \end{pmatrix}&
  \VV &=
  \begin{pmatrix}
    v_1 \\ v_2\\
  \end{pmatrix}&
  N_f &= 
  \begin{pmatrix}
    1 & 0 \\
    0 & 0 \\
    0 & 1 \\
    1 & 0
  \end{pmatrix}&
  N_r &= 
  \begin{pmatrix}
    0 & 0 \\
    0 & 1 \\
    1 & 0 \\
    0 & 1
  \end{pmatrix}&
  N &= 
  \begin{pmatrix}
    -1 & 0 \\
    0 & 1 \\
    1 & -1 \\
    -1 & 1
  \end{pmatrix} 
\end{xalignat}
Substituting into Equation \eqref{eq:V} gives:
\begin{xalignat}{2}
  v_1 &= \kappa_1 \lb K_A K_E x_A x_E - K_C x_C \rb&
  v_2 &= \kappa_2 \lb K_C x_C -  K_B K_E x_B x_E\rb
\end{xalignat}
and substituting into Equation \eqref{eq:Xdot} gives:
\begin{xalignat}{3}
  \dx_A &= - v_1 &
  \dx_B &=  v_2 & 
  \dx_C &= -\dx_E = \lb v_1 - v_2 \rb 
\end{xalignat}

\subsection{Chemostats}\label{sec:chemostats}
As discussed by \citet{PolEsp14}, the notion of a \emph{chemostat} is
useful in creating an open system from a closed system;
a similar approach is used by \citet{QiaBea05} who use the phrase
``concentration clamping''.
The chemostat has three interpretations:
\begin{enumerate}
\item one or more species is fixed to give a constant concentration
  \citep{GawCurCra15}; this implies that an appropriate external
  flow is applied to balance the internal flow of the species.
\item an ideal feedback controller is applied to species to be fixed
  with setpoint as the fixed concentration and control signal an
  external flow.
\item as a \C component with a fixed state.
\end{enumerate}

Define $\II^{cs}$ as the set containing the indices of the species
corresponding to the chemostats. Then the $n_X\times n_X$ diagonal
matrices $I^{cs}$ and $I^{cd}$ are defined as:
\begin{xalignat}{2}
  \label{eq:I_cs}
  I^{cs}_{ii} &= 
  \begin{cases}
    1 & \text{if $i \in \II^{cs}$}\\
    0 & \text{if $i \not\in \II^{cs}$}
  \end{cases}&
  I^{cd}_{ii} &= 
  \begin{cases}
    0 & \text{if $i \in \II^{cs}$}\\
    1 & \text{if $i \not\in \II^{cs}$}
  \end{cases}
\end{xalignat}                  
It follows that $I_{X} = I^{cs} + I^{cd}$ where $I_X$ is the $n_X\times n_X$
unit matrix.  The stoichiometric matrix $N$ can then be expressed as
the sum of two matrices: the \emph{chemostatic} stoichiometric matrix
$N^{cs}$ and the \emph{chemodynamic} stoichiometric matrix$N^{cd}$ as
\begin{align}
  N &= N^{cs} + N^{cd} \\ 
\text{where } N^{cs} &= I^{cs}N 
\text{ and } N^{cd} = I^{cd}N \label{eq:N^cd}
\end{align}
Note that $N^{cd}$ is the same as $N$ except that the \emph{rows}
corresponding to the chemostat variables are set to zero. 
The stoichiometric properties of $N^{cd}$, rather than $N$, determine
system properties when chemostats are present. 
%
When chemostats are used, the state equation \eqref{eq:Xdot} is
replaced by:
\begin{equation}
  \label{eq:dX_cd}
  \dot{X} = N^{cd} V = NV - V^{cs} \text{ where } V^{cs} = N^{cs} V 
\end{equation}
and thus the fixed states are held constant by the external flow
flow $V^s = - V^{cs} = - N^{cs}V$ acting at the \C components. Thus
the closed-system bond graph  of Figure \ref{subfig:Closed_bg} is
replaced by the open-system bond graph  of Figure \ref{subfig:Open_bg}
where the external flows $V^s$ have been added.

\subsection{Flowstats}\label{sec:flowstats}
In addition to ``concentration clamping'' (identified with chemostats
in \S~\ref{sec:chemostats}), \citet{QiaBea05} also use ``boundary flux
injection'' to convert closed to open systems. Here we ``fix'' flows
though \Re components to create flowstats.
Although \citet{PolEsp14} ``focus on chemostats for thermodynamic
modelling'' and note that chemostats can be used to to create fixed
currents,  it is argued that flowstats provide a useful complement to
chemostats.

In a similar way to \S~\ref{sec:chemostats}, define $\II^{fs}$ as the
set containing the indices of the \emph{reactions} corresponding to the
flowstats. Then the $n_V\times n_V$ diagonal matrices $I^{fs}$ and
$I^{fd}$ are defined as:
\begin{xalignat}{2}
  \label{eq:I_fs}
  I^{fs}_{ii} &= 
  \begin{cases}
    1 & \text{if $i \in \II^{fs}$}\\
    0 & \text{if $i \not\in \II^{fs}$}
  \end{cases}&
  I^{fd}_{ii} &= 
  \begin{cases}
    0 & \text{if $i \in \II^{fs}$}\\
    1 & \text{if $i \not\in \II^{fs}$}
  \end{cases}
\end{xalignat}                  
It follows that $I_{V} = I^{fs} + I^{fd}$ where $I_{V}$ is the $n_V\times n_V$
unit matrix. Thus the flows $V$ are replaced by $V^{cd}$ where:
\begin{equation}
  V^{cd} = I^{fd}V + I^{fs}V^{fs} 
\end{equation}
Assuming that chemostats are also present, Equation \ref{eq:dX_cd} is
replaced by
\begin{equation}
  \label{eq:dX_d}
  \dot{X} = N^{cd} V^{cd} = N^{cd} \lb I^{fd}V + I^{fs}V^{fs}\rb 
\end{equation}
If $V^{fs} \ne 0$, the stoichiometric properties of $N^{cd}$ (ie
determined by the chemostats) determine
system properties. However if $V^{fs} = 0$ then the stoichiometric
properties of
\begin{equation}\label{eq:N^d}
  N^d =  N^{cd}  I^{fd} = I^{cd}NI^{fd}
\end{equation}
(that is both chemostats and flowstats) determine system properties.
Note that $N^{d}$ is the same as $N^{cd}$ except that the \emph{columns}
corresponding to the flowstat variables are set to zero.

Figure \ref{subfig:Open_bd_simp} is the block diagram corresponding to
the open system bond graph of Figure \ref{subfig:Open_bg}. It differs
from Figure \ref{subfig:Closed_bg} in that $N$ of Equation
\eqref{eq:mmu} is replaced by $N^{cd}$ of Equation \eqref{eq:dX_cd} to
reflect the fact that the chemostat states are not affected by $V$ and
thus correspond to the zero rows of $N^{cd}$. Moreover, the matrices
$I^{fd}$ and $I^{fds}$, and the flows $V^{fs}$ are added to reflect
the effect of the flowstats.


\subsection{Reduced-order equations}
\label{sec:RO}
\begin{figure}[htbp]
  \centering
  \SubFig{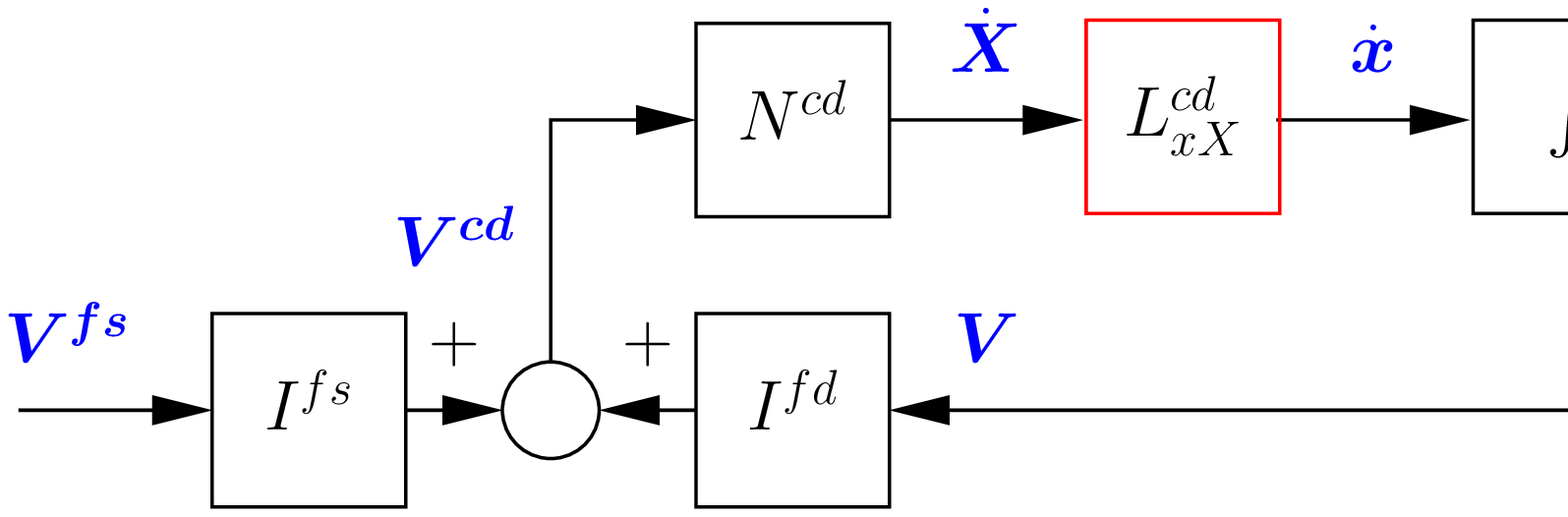}{Reduced-order system}{0.9}
  \SubFig{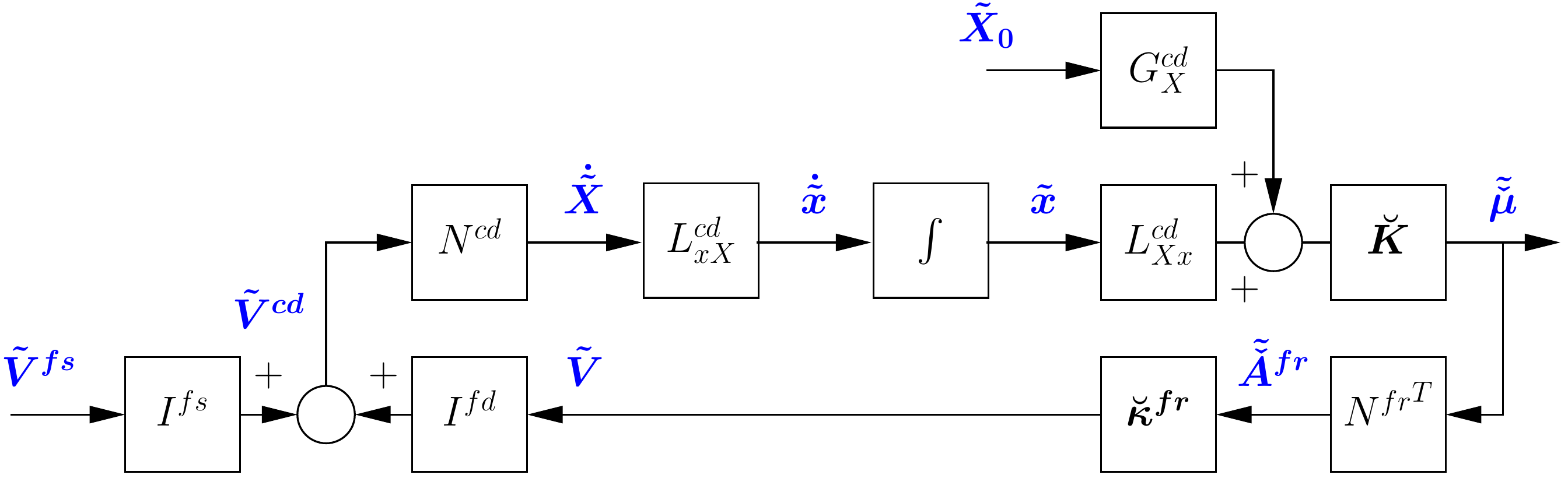}{Linearised Reduced-order system}{0.9}
  \caption{Reduced-order system block diagrams.
    (a) Reduced-order system. 
    (b) Linearised system corresponding to (a).
}\label{fig:RO}
\end{figure}
As discussed by number of authors
\citep{Sau09,Ing13}, 
the presence of conserved
moieties leads to potential numerical difficulties with the solution of
Equation \eqref{eq:Xdot}. As chemostats introduce further conserved
moieties it is important to resolve this issue. The following outline
uses the notation and approach of \citet[\S 3(c)]{GawCra14}.

Defining $\Gcd$ as the left null-space matrix of $N^{cd}$ it follows that:
\begin{equation}
  \label{eq:G_cm}
  \Gcd\dot{\XX} = \Gcd \Ncd \VV = 0
\end{equation}
Hence each  of the $n_G$ rows of $\Gcd$ defines an algebraic relationship between the
states contained in $\XX$. Thus the number of \emph{independent} states $n_x$
is given in terms of the \emph{total} number of states $n_X$ by:
\begin{equation}
  \label{eq:n_x}
  n_x = n_X - n_G
\end{equation}
The derivative of the independent states $\xx$ is given in terms of the
derivative of state $\XX$ by
the $n_x \times n_X$ transformation matrix $\Lcd_{xX}$
\begin{equation}
  \label{eq:dxdX}
  \dot{\xx} = \Lcd_{xX} \dot{\XX}
\end{equation}
Similarly
\begin{equation}
  \label{eq:dXdx}
  \dot{\XX} = \Lcd_{Xx} \dot{\xx}
\end{equation}
where $\Lcd_{Xx}$ is an $n_X \times n_x$ matrix.
Integrating equation \eqref{eq:dXdx}, 
\begin{align}
  \XX  &= \Lcd_{Xx} \xx + \XX_0 - \Lcd_{Xx}\xx_0 = \Lcd_{Xx} \xx + \Gcd_X \XX_0\label{eq:Xx}\\
  \text{where } \Gcd_X &= I_{n_X \times n_X} - \Lcd_{Xx}\Lcd_{xX}\label{eq:G_X}
\end{align}
and $\xx_0$ and $\XX_0$ are the values of $\xx$ and $\XX$ at time
$t=0$.

Figure \ref{subfig:Open_bd_ro_simp} corresponds to the open system bond graph and
block diagram of Figure \ref{fig:Closed}, but the reduced-order
equations \eqref{eq:dxdX} and ~\eqref{eq:Xx} have been incorporated.
The block $\Lcd_{xX}$ contracts the state dimension from $n_X$ to
$n_x$ and the block $\Lcd_{Xx}$ expands it again. The initial
condition term $G_X \XX(0)$ becomes an exogenous signal analogous to
the setpoint term of feedback control; note that this includes the
states of all of the chemostats.

\subsection{Examples}
\label{sec:ex_stats}
The simple reaction $A \reacu{1} B$ of Figure
\ref{subfig:CRCRe_abg} has a single conserved moiety represented by
\begin{equation}
  \label{eq:CRCRe_cm}
  x_A + x_B = x_{AB}
\end{equation}
where $x_{AB}$ is a constant.
One possibility is
\begin{xalignat}{4}
\label{eq:ABEC_ro}
  \xx &=
  \begin{pmatrix}
    x_A
  \end{pmatrix}&
  L_{xX} &=
  \begin{pmatrix}
    1 & 0 
  \end{pmatrix}&
  L_{Xx} &=
  \begin{pmatrix}
    1\\-1
  \end{pmatrix}&
  G_X &= 
  \begin{pmatrix}
    0&0\\1 & 1
  \end{pmatrix}
\end{xalignat}

The enzyme-catalysed reaction $A+E \reacu{1} C
\reacu{2} B+E$ of Figure \ref{subfig:ABEC_abg} has a number of
possible representations depending on which \C components are
chemostats and which \Re components are flowstats. Two of these are
examined here.

Firstly, consider the case where both \BC{A} and \BC{B} are chemostats
and \BRe{r0} is a flowstat with zero flow. The relevant stoichiometric
matrix is thus $N^d$ of Equation \eqref{eq:N^d} that determines system
properties and
\begin{xalignat}{2}
  N^{d} &=
  \begin{pmatrix}
    0&0&0\\
    0&0&0\\
    0&1&-1\\
    0&-1&1
  \end{pmatrix} &
  G^{d} &=
  \begin{pmatrix}
    1&0&0&0\\
    0&1&0&0\\
    0&0&1&1
  \end{pmatrix}
\end{xalignat}
$G^d$ has three rows corresponding to the three conserved moieties
$x_A$, $x_B$ and $x_C + x_E$. The first correspond to the two
chemostats, the third to the well-known conserved moiety for
enzyme-catalysed reactions: the total enzyme amount is
conserved. There is only one independent state which is chosen as
$x_C$. With this choice:
\begin{xalignat}{3}
\label{eq:ABEC_d_ro}
L_{xX} &=
\begin{pmatrix}
  0&0&1&0
\end{pmatrix} &
L_{Xx} &=
\begin{pmatrix}
  0\\
  0\\
  1\\
  -1
\end{pmatrix}&
G_X &=
\begin{pmatrix}
  1&0&0&0\\
  0&1&0&0\\
  0&0&0&0\\
  0&0&1&1
\end{pmatrix}
\end{xalignat}

Secondly, consider the case where both \BC{A} and \BC{B} are chemostats
and \BRe{r0} is a flowstat with non-zero flow. The relevant stoichiometric
matrix is thus $N^cd$ of Equation \eqref{eq:N^cd} that determines system
properties and
\begin{xalignat}{2}
  N^{cd} &=
  \begin{pmatrix}
    0&0&0\\
    0&0&0\\
    0&1&-1\\
    1&-1&1
  \end{pmatrix}&
  G^{cd} &=
  \begin{pmatrix}
    1&0&0&0\\
    0&1&0&0
  \end{pmatrix}
\end{xalignat}
The effect of the variable flowstat is to remove the third conserved
moiety leaving only the chemostat states $x_A$ and $x_B$.
There are now two independent states $x_C$ and $x_E$. This gives:
\begin{xalignat}{3}
  \label{eq:ABEC_cd_ro}
  L_{xX} &=
  \begin{pmatrix}
    0&0&1&0\\
    0&0&0&1
  \end{pmatrix}
  &
  L_{Xx} &=
  \begin{pmatrix}
    0&0\\
    0&0\\
    1&0\\
    0&1
  \end{pmatrix}
  &
  G_X &=
  \begin{pmatrix}
    1&0&0&0\\
    0&1&0&0\\
    0&0&0&0\\
    0&0&0&0
  \end{pmatrix}
\end{xalignat}

\section{Linearisation}
\label{sec:linearisation}
As discussed in the Introduction, linearisation of non-linear systems
is a standard technique in control engineering.  In
\S~\ref{sec:feedback} of this paper, linearisation is used to analyse
the properties of modules. 

Assuming that the system reaches a steady-state $\sss{\XX}$, that is
$\dot{\XX}=0$ when $\XX=\sss{\XX}$, the system can be linearised about that
steady state by introducing  \emph{perturbation variables} $\ttt{\XX}$
so that $\XX = \ttt{\XX} + \sss{\XX}$. These can be defined for each
relevant variable; for example:
\begin{xalignat}{5}
  \label{eq:pert}
  x &= \xs + \ttt{x} &
  \mu &= \sss{\mu} + \ttt{\mu} &
  A^f &= \sss{A^f} + \ttt{A^f}  &
  A^r &= \sss{A^r} + \ttt{A^r}  &
  v &= \sss{v} + \ttt{v} 
\end{xalignat}
If the perturbation is small, each variable can be approximated using a
first-order Taylor series; thus, for example $\ttt{\mu} \approx
\frac{\partial \mu}{\partial x} \ttt{x}$.

\subsection{Component linearisation: \C and \Re}
\label{sec:comp-linear}

The non-linear \C component is defined by the equations
\eqref{eq:C_chem_norm}. In particular, for substance $A$:
\begin{xalignat}{2}
  \label{eq:C_nlin}
  \nmu_A &= \ln K_Ax_A &
  \dot{x}_A &= v_A
\end{xalignat}
Using the perturbation approach, it follows that the linearised \C
component is defined by the equations:
\begin{xalignat}{3}
  \label{eq:C_lin}
  \ttt{\nmu}_A &= \Kt_A \ttt{x}_A& 
  \Kt_A &= \pd{\nmu_A}{x_A} = \frac{1}{\xs_A}&
  \dot{\ttt{x}}_A &= \ttt{v}_A
\end{xalignat}

The non-linear \Re component representing the $ith$ reaction
\eqref{eq:react} is defined by equations \eqref{eq:v_exp},
in particular:
\begin{xalignat}{3}
  \label{eq:R_nlin}
  v_i &= \kappa_i \lb v_0^+ - v_0^-\rb &
  v_0^+ &= e^{\nA^f_i}&
  v_0^- &= e^{\nA^r_i}
\end{xalignat}
Hence the linearised \Re component is defined by the equations:
\begin{xalignat}{3}
  \label{eq:R_lin}
  \vt_i &= \kt^f_i \ttt{\nA}^f_i - \kt^r_i \ttt{\nA}^r_i& 
  \kt^f_i &= \pd{v_i}{\nA^f_i} = \kappa_i {\sss{v}_0^+}&
  \kt^r_i &= -\pd{v_i}{\nA^r_i} = \kappa_i \sss{v}_0^-
\end{xalignat}

\subsection{Linearised system equations}
\label{sec:line-syst-equat}
Section \ref{sec:comp-linear} shows how the bond graph
\emph{components} are linearised; essentially the non-linear $\exp$
and $\ln$ functions are replaced by linear gains dependent on the
steady-state flows and steady-state states respectively.  The $n_X$
constants $\Kt_A, \Kt_B, \dots$ of the linearised \C components, and
the $n_V$ constants $\kt^f_1, \kt^f_2, \dots$ and $\kt^r_1, \kt^r_2,
\dots$ are collected into column vectors:
\begin{xalignat}{3}
  \label{eq:Kt}
  \Kt &=
  \begin{pmatrix}
    \Kt_A\\\Kt_B\\\vdots
  \end{pmatrix}&
  \kt^f &=
  \begin{pmatrix}
    \kt^f_1\\\kt^f_2\\\vdots
  \end{pmatrix}&
  \kt^r &=
  \begin{pmatrix}
    \kt^r_1\\\kt^r_2\\\vdots
  \end{pmatrix}
\end{xalignat}
Figure 
\ref{subfig:Open_bd_ro_simp_lin} shows the block diagram corresponding
to the linearisation of the reduced-order system depicted in Figure
\ref{subfig:Open_bd_ro_simp} where:
\begin{xalignat}{3}
  \lin{\KK} &= \diag{\lin{K}} &
  \lin{\kkappa} &= \diag{\lin{\kappa}^{fr}} &
\text{where } \lin{\kappa}^{fr} &=
\begin{pmatrix}
  \lin{\kappa}^f\\ \lin{\kappa}^r
\end{pmatrix}
\end{xalignat}

The block diagram of the linearised version of the full system of Figure
\ref{subfig:Open_bd_simp} gives the following linear state-space
equations:
\begin{xalignat}{3}
  \label{eq:dX_lin}
  \ttt{\VV} &= \CC \ttt{\XX} &
  \ttt{\VV}^{cd} &= I^{fd} \ttt{\VV} + I^{fs} \ttt{\VV}^{fs}&
  \dot{\ttt{\XX}} &= \AA \ttt{\XX} + \BB \ttt{\VV}^{fs}
\end{xalignat}
where
\begin{xalignat}{3}\label{eq:ABC_lin}
  \AA &= N^{cd}I^{fd}\CC = N^{d}\CC&
  \BB &= N^{cd}I^{fs}&
  \CC &= \lin{\kkappa}^{fr} \Nfr^T \lin{\KK}
\end{xalignat}

The block diagram of the reduced system in Figure
\ref{subfig:Open_bd_ro_simp_lin} gives the following linear
state-space equations:
\begin{xalignat}{2}
  \label{eq:dx_lin}
  \ttt{\VV} &= \CCC \ttt{\xx} + \DDD \ttt{\XX}_0&
  \dot{\ttt{\xx}} &= \AAA \ttt{\xx} + \BBB_v \ttt{\VV}^{fs} +  \BBB_x \ttt{\XX}_0
\end{xalignat}
where
\begin{xalignat}{5}\label{eq:abc_lin}
  \AAA &= L^{cd}_{xX} \AA L^{cd}_{Xx}&
  \BBB_v &= L^{cd}_{xX} \BB&
  \BBB_x &= L^{cd}_{xX} \AA G^{cd}_X&
  \CCC &= \CC L^{cd}_{Xx} &
  \DDD &= \CC G^{cd}_X
\end{xalignat}

Equations \eqref{eq:dx_lin} can be written more compactly as:
\begin{align}
  \label{eq:dx_lin_compact}
  \ttt{\VV} &= \CCC \ttt{\xx} + \DDD \ttt{\UU}\\
  \dot{\ttt{\xx}} &= \AAA \ttt{\xx} + \BBB \ttt{\UU}\\
  \text{where } \UU &=
  \begin{pmatrix}
    \ttt{\XX}_0 \\ \ttt{\VV}^{fs}
  \end{pmatrix}\\
  \BBB &=
  \begin{pmatrix}
    \BBB_x & \BBB_v
  \end{pmatrix}\\
  \text{and }
  \DDD &=
  \begin{pmatrix}
    \DDD_x & 0_{n_x \times n_V}
  \end{pmatrix}
\end{align}
where $0_{n_x \times n_V}$ is the zero matrix with indicated
dimensions.

Because the state-space systems \eqref{eq:dX_lin} and
\eqref{eq:dx_lin} are linear, they can also be represented as transfer
functions in the Laplace variable $s$. In particular, the
reduced-order system \eqref{eq:dx_lin} has the transfer function
$\GG(s)$ given by
\begin{align}
  \GG(s) &= \CCC \lb sI_{n_x \times n_x} - \AAA \rb^{-1} \BBB + \DDD  \label{eq:TF}\\
\text{where } \VV(s) &= \GG(s) \UU(s)\notag
\end{align}
 $I_{n_x \times n_x}$ is the unit matrix with indicated dimensions.

\subsection{Examples}
\label{sec:lin_examples}
The simple reaction $A \reacu{1} B$ of Figure \ref{subfig:CRCRe_abg}
has a flow given by \eqref{eq:simple_v}. As both state derivatives are proportional to $v$, it
follows that the steady-state is defined by: $v = \kappa \lb K_A x_A - K_B
x_A \rb = 0$. As noted in Equation \eqref{eq:CRCRe_cm}, $x_A + x_B =
x_{AB}$ where $x_{AB}$ is a constant. It follows that the steady-state
values of $x_A$ and $x_B$ are
\begin{xalignat}{2}\label{eq:simple_ss}
  \sss{x}_A &= \frac{K_B}{K_A+K_B}x_{AB}&
  \sss{x}_B &= \frac{K_A}{K_A+K_B}x_{AB}
\end{xalignat}
From Equation \eqref{eq:simple_v_0} $v^+_0=K_A x_A$ and $v^-_0=K_B
x_B$. Hence
\begin{equation}
  \sss{v}^+_0 = \sss{v}^-_0 = \frac{K_AK_B}{K_A+K_B}x_{AB}
\end{equation}
Using the formulae \eqref{eq:C_lin} and \eqref{eq:R_lin}, it follows
that the coefficients of the linearised \C components are:
\begin{xalignat}{2}
  \label{eq:simple_lin_c}
  \Kt_A &= \frac{1}{\xs_A} = \frac{K_A+K_B}{K_B\xs_{AB}}&
  \Kt_B &= \frac{1}{\xs_B} = \frac{K_A+K_B}{K_A\xs_{AB}}
\end{xalignat}
and that the coefficients of the linearised \Re component are:
\begin{equation}
  \label{eq:simple_lin_re}
  \kt^f = \kt^r = \kappa \frac{K_AK_B}{K_A+K_B}x_{AB}
\end{equation}
Hence the linearised equations for the flow are:
\begin{equation}
  \ttt{v} = \kt^f\Kt_A \ttt{x}_A - \kt^r\Kt_B \ttt{x}_B 
  = \kappa \lb K_A \ttt{x}_A - K_B \ttt{x}_B\rb
\end{equation}
As expected, the linearisation of a linear equation is the same as the
linear equation.

\section{Modularity, Retroactivity and Feedback}
\label{sec:feedback}
\begin{figure}[htbp]
  \centering
  \SubFig{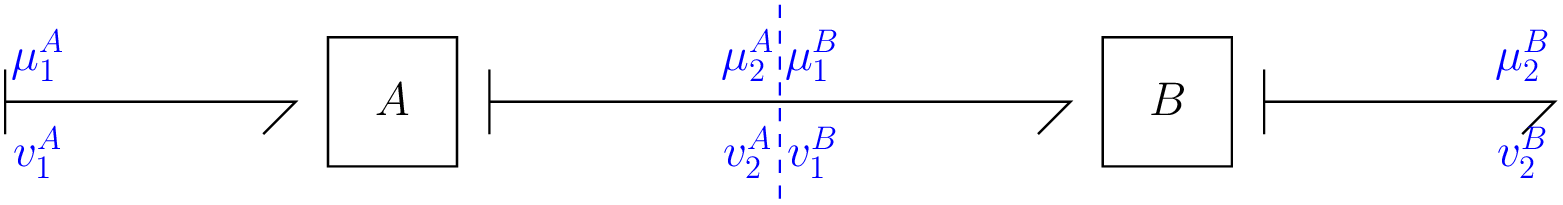}{Series connection: bond graph}{0.7}
  \SubFig{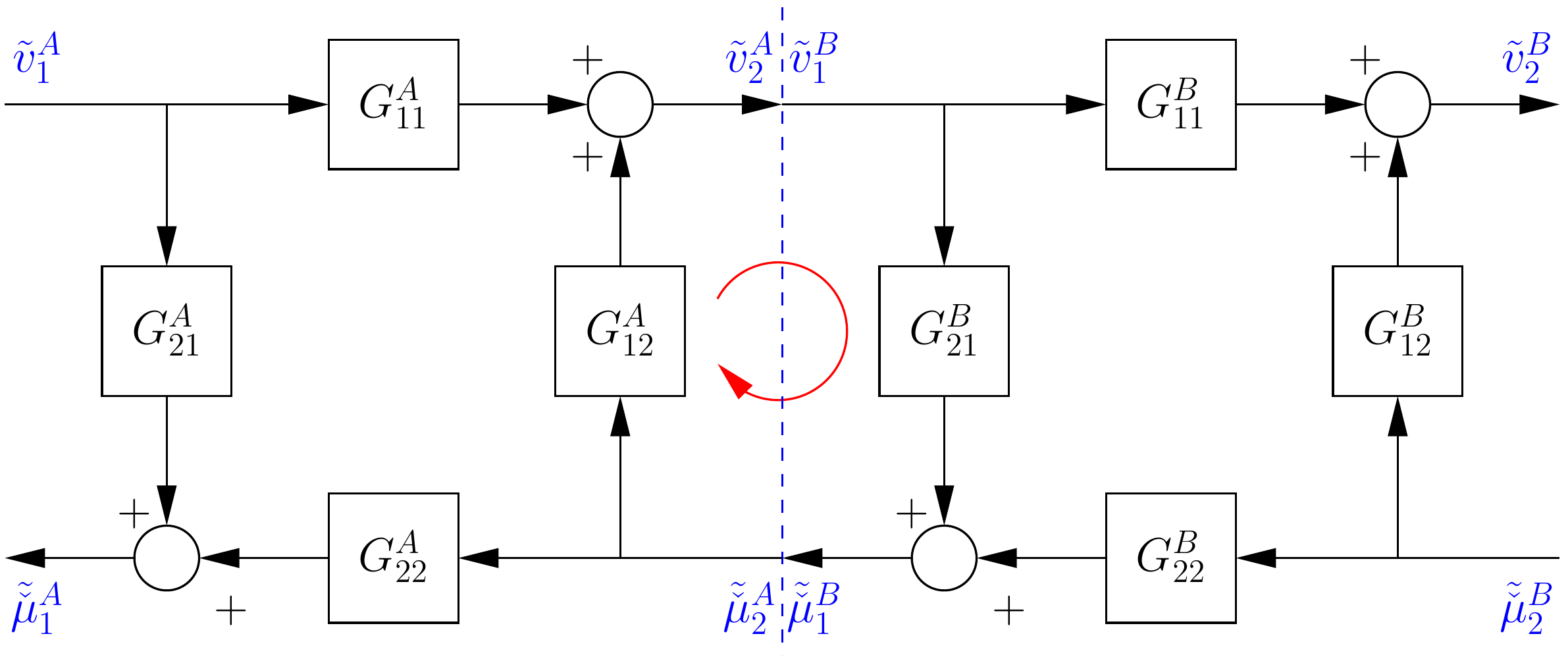}{Series connection: linearised block diagram}{0.7}
  \caption{Modularity, Retroactivity and Feedback}
  \label{fig:modularity}
\end{figure}
Modularity provides one approach to understanding the complex systems
associated with biochemical systems
\citep{HarHop99,Lau00,CseDoy02,BruWesHoe02,BruSnoWes08,SzaPerSte10}.
However, as discussed by \citet{KalSte12} there are many possible
concepts of modularity.
These include structure deduced from the stoichiometric matrix
\citep{SchKhoWes00,PooSebPid07};
modular construction of \emph{in silico} models \citep{NeaCooSmi14};
and modular structure designed to minimise the
\emph{retroactivity} between modules \citep{KalSte12,Vec13,VecMur14}.

This paper focuses on two overlapping, but conceptually different
concepts of modularity:
\begin{description}
\item [Computation modularity:]  modules retain physically correct
  results when connected together to form a system
\item[Behavioural modularity:] modules retain their behaviour (such as
  ultrasensitivity) when connected together.
\end{description}

\citet{GawCurCra15}
have shown that bond graphs provide an effective foundation for
modular construction of computer models of biochemical systems.  This
paper focuses on the second interpretation of modularity and shows
that bond graphs provide a natural interpretation of inter-module
\emph{retroactivity}~\citep{SaeKreCon04,SaeKreGil05,VecNinSon08,Son11,JayVec11,Vec13,VecMur14}.
%
%
Retroactivity has been illustrated experimentally in the context of ``signalling properties of a
covalent modification cycle'' \citep{VenJiaWas10}, ``load-induced modulation of signal
transduction networks'' \citep{JiaVenSon11} and the  ``temporal dynamics of
gene transcription'' \citep{JayNigVec13}.
%
Retroactivity can be removed using ``insulation''
\citet{VecSon09,Son11,VecMur14};  however, this may come at an
energetic cost \citep{BarSon12}.

As discussed in the Introduction, feedback is another concept crucial
to the understanding of complex systems.  
\citet{Kho00}, \citet{BriFel00},
\citet{AstLau01}, \citet{KolCalGil05}, \citet{HorBinBru05} and
\citet{SauIng07} investigate
the  feedback in the context of MAPK cascades.
As will be shown in this paper, retroactivity and feedback are closely
related concepts.
As will be seen, feedback arises in a number of ways including:
\begin{description}
\item[Intrinsic feedback] due to the interaction of reactions and
  species within and between modules
\item[Conserved moieties] implicitly generate feedback loops
\item[Feedback inhibition] explicitly uses negative feedback.
\end{description}

As discussed in \S~\ref{sec:linearisation}, linearisation of a
non-linear system allows a wide range of control engineering
techniques to be applied. In this section linearisation is used to
investigate behavioural modularity using transfer functions and
frequency-domain methods.

Figure \ref{subfig:BG_series} shows the series interconnection of two
bond graph modules labelled $A$ and $B$. In this example, each module
has two ports labelled $1$ and $2$ and the modules are interconnected
to form a composite module $AB$ with two ports.
To create a block diagram from a bond graph, the concept of causality
is required. This concept is discussed in detail in the textbooks
\citep{Wel79,GawSmi96,MukKarSam06,KarMarRos12}, but here it suffices
to know that causality determines which variable on a bond impinging
on a system is the input, and which the output.
For example, in this case
the causality is such that flow $v$ is the input (and effort $\mu$ the output)
on port 1 and that effort $\mu$ is the input (and flow $v$ the output) on port 2.
%

As discussed by \citet{GawCurCra15}, the bond graph approach can be
used to build arbitrarily complex systems out of such modules.
However, to delve more deeply into the power of the bond graph
approach and to understand how modules interact, it is instructive to
look at the block diagram equivalents \emph{following linearisation}
as discussed in \S~\ref{sec:linearisation}.
With the assumed causality, each module can be
represented by four transfer functions $G_{11}$,$G_{12}$,$G_{21}$ and
$G_{22}$ which can be combined into a $2\times2$ matrix:
\begin{equation}\label{eq:module_tf}
  \begin{pmatrix}
    \vt_2\\\ttt{\nmu}_1
  \end{pmatrix} =
  \begin{pmatrix}
    G_{11} & G_{12}\\
    G_{21} & G_{22}
  \end{pmatrix}
  \begin{pmatrix}
    \vt_1\\\ttt{\nmu}_2
  \end{pmatrix}
\end{equation}
Using the superscripts $A$ and $B$ to refer to the two modules, the
four transfer functions of Equation \eqref{eq:module_tf} can be
represented for each of the interconnected modules as Figure
\ref{subfig:BD_series}. Connecting port 2 of $A$ to port $1$ of $B$ in
Figure \ref{subfig:BG_series} is equivalent to connecting the
corresponding signals in Figure \ref{subfig:BD_series}:
\begin{equation}\label{eq:module_connect}
  \vt^A_2 = \vt^B_1 \text{ and } \ttt{\nmu}^A_2 = \ttt{\nmu}^B_1 
\end{equation}
This connection induces a feedback loop involving $G^{A}_{12}$ and
$G^{B}_{21}$ thus the properties of the composite system are dependent
on the loop gain $L_I$ of this feedback loop.

In particular, using Equations \eqref{eq:module_tf} for $A$ and $B$
and substituting \eqref{eq:module_connect} gives the transfer function
$G^{AB}$ for the composite module as
\begin{align}
  G^{AB}_{11} &= \frac{G^{A}_{11} G^{B}_{11}}{1+L_I}\label{eq:Gvv}\\
  G^{AB}_{12} &= G^{B}_{12} + \frac{G^{B}_{11} G^{A}_{12}G^{B}_{22}}{1+L_I}\label{eq:Gev}\\
  G^{AB}_{21} &= G^{A}_{21} + \frac{G^{A}_{22} G^{B}_{21}G^{A}_{11}}{1+L_I}\label{eq:Gve}\\
  G^{AB}_{22} &= \frac{G^{B}_{22}
    G^{A}_{22}}{1+L_I}   \label{eq:Gee}\\
  L_I &= -G^{A}_{12}G^{B}_{21} \label{eq:L_I}
\end{align}
$L_I$ will be called the \emph{interaction loop-gain}.
In linear systems, feedback shifts system poles and therefore changes
the behaviour of the interacting systems. In particular, each of the
transfer functions $G^{AB}_{ij}$ of equations
\eqref{eq:Gvv}~--~\eqref{eq:Gee} is modified by the interaction
loop-gain. Thus the feedback loop comprising $G^{A}_{12}$ and
$G^{B}_{21}$ is the source of behaviour alteration when two modules
are connected.  It follows that approximate behavioural modularity is
achieved by making the interaction loop-gain as small as possible.
Indeed, in the special case that $G^{A}_{12} = G^{B}_{21} = 0$ and so $L_I=0$ then:
\begin{xalignat}{5} \label{eq:L_0}
  G^{AB}_{11} &= G^{A}_{11} G^{B}_{11},&
  G^{AB}_{21} &= G^{A}_{21} &
  G^{AB}_{12} &= G^{B}_{12} &
  G^{AB}_{22} &= G^{B}_{22} G^{A}_{22}
\end{xalignat}

\subsection{Example module: Simple reaction}
\label{sec:species-reaction}
\begin{figure}[htbp]
  \centering
  \SubFig{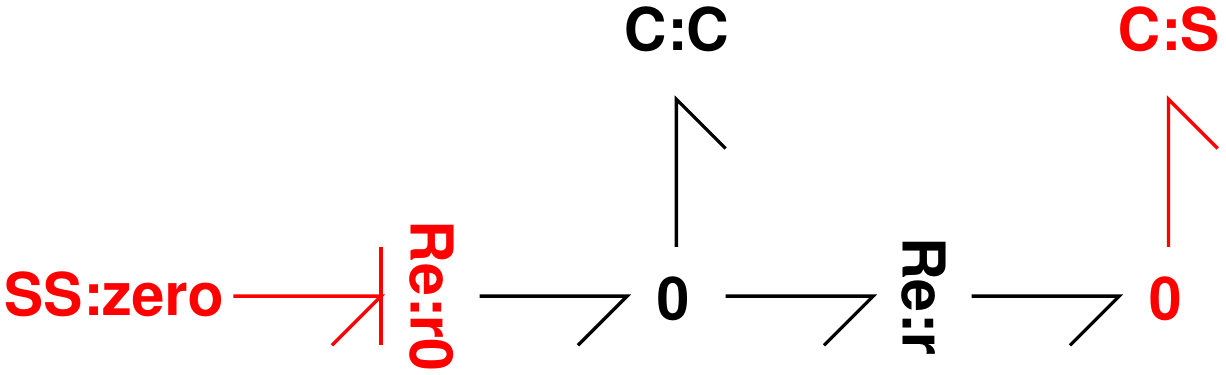}{With flowstat and chemostat}{0.45}
  \SubFig{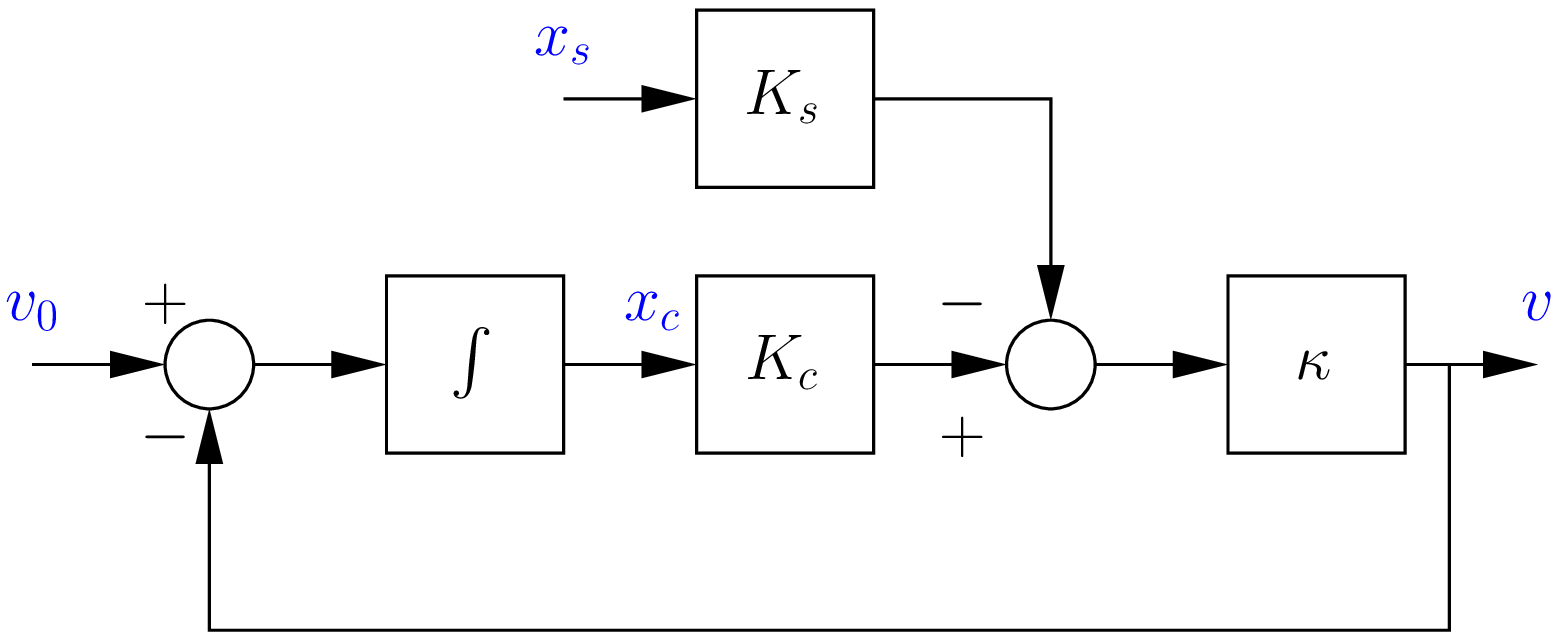}{Block diagram}{0.45}
  \SubFig{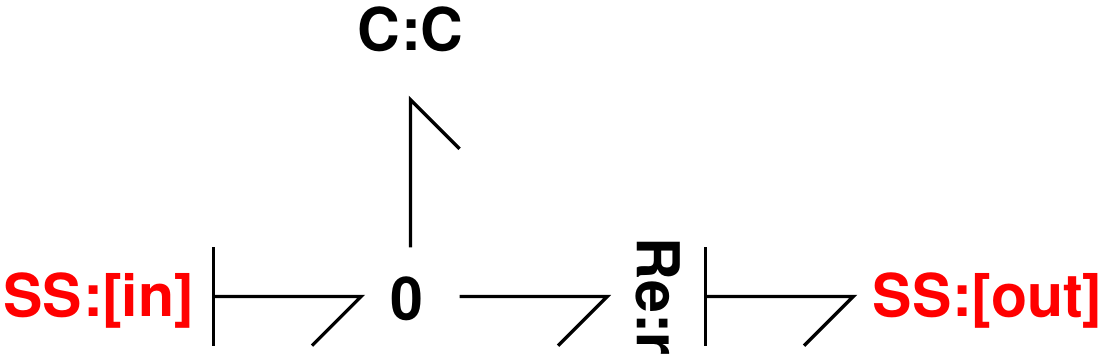}{With ports}{0.45}
  \SubFig{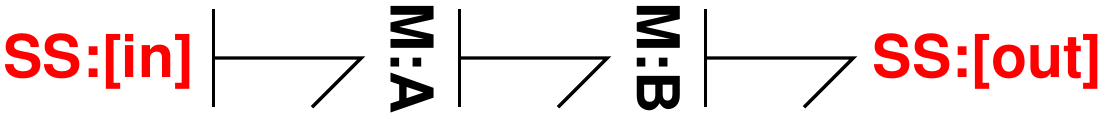}{Example series connection}{0.45}
  \caption{Example module: simple reaction}
\label{fig:species-reaction}
\end{figure}
Figure \ref{subfig:cM_abg} shows a simple reaction system
comprising a species represented by the \C component \BC{C} and the
reaction component \BRe{r}. This closed system is converted to an open
system by appending a flowstat \BRe{r0} with flow $v_0$ and a
chemostat \BC{S} with state $x_s$. This system is linear and the
reaction flow (though \BRe{r}) is given by:
\begin{equation}
  \label{eq:SR_v}
  v = \kappa \lb K_c x_c - K_s x_s \rb
\end{equation}
and the rate of change of $x_c$ is
\begin{equation}
  \label{eq:SR_x}
  \dot{x}_c = v_0 -v
\end{equation}
Equations \eqref{eq:SR_v} and \eqref{eq:SR_x} can be visualised using
the block diagram of Figure \ref{subfig:M_bd} which clearly shows the
implicit feedback structure with loop gain
\begin{equation}
  \label{eq:SR_L}
  L = \frac{\kappa K_c}{s}
\end{equation}




It follows from the block diagram of Figure \ref{subfig:M_bd} that:
\begin{equation}
  \label{eq:SR_G}
  \begin{pmatrix}
    v\\x_c
  \end{pmatrix} =
  \frac{1}{1+L}
  \begin{pmatrix}
    L & \kappa K_s\\
    \frac{1}{s} &  \frac{\kappa K_s}{s}
  \end{pmatrix}
  \begin{pmatrix}
    v_0\\x_s
  \end{pmatrix} =
  \frac{1}{s + \kappa K_c}
  \begin{pmatrix}
    \kappa K_c & -s\kappa K_s\\
    1 & \kappa K_s
  \end{pmatrix}
  \begin{pmatrix}
    v_0\\x_s
  \end{pmatrix}
\end{equation}

In the particular case that $\kappa = K_c = K_s = 1$
\begin{xalignat}{4}
  \label{eq:M_tf}
  G_{11} &= G_{22} = G_{21} = \frac{1}{s+1} &
  G_{12} &= \frac{-s}{s+1} 
\end{xalignat}

If two identical copies of this module are placed in series as in 
Figure \ref{subfig:MM_abg}, 
\begin{xalignat}{2}
    L_I &= \frac{s}{(s+1)^2}&
  \frac{1}{1+L_I} &= \frac{(s+1)^2 }{s^2+3s+1}
\end{xalignat}
and the resulting overall transfer function is:
\begin{xalignat}{3}
  G^{AB}_{11} &= G_{22} = \frac{1}{s^2+3s+1}&
  G^{AB}_{21} &= \frac{-s(s+2)}{s^2+3s+1}&
  G^{AB}_{12} &= \frac{s+2}{s^2+3s+1}
\end{xalignat}
The isolated modules each have a single pole at $s=-1$; the series
modules has a pole at $s=-0.38$ and at $s=-2.62$. This shift in pole
location is due to non-zero interaction loop-gain $L_I$ \eqref{eq:L_I}.
%

Such reaction systems are often incorrectly modelled using an
irreversible reaction where the flow is independent of $\mu_2$. This
would imply that $G_{12}=G_{22}=L=0$ and thus the overall transfer
function would be
\begin{xalignat}{3}
  G^{AB}_{11} &= \frac{1}{(s+1)^2} = \frac{1}{s^2 + 2s + 1}&
  G^{AB}_{21} &= \frac{-s}{s+1}&
  G^{AB}_{12} &= G^{AB}_{22} = 0
\end{xalignat}
This thermodynamically incorrect system has zero retroactivity. As
will be shown in the sequel, approximate irreversibility, and thus
approximate zero retroactivity, can be achieved but at the metabolic
cost of using a power supply such as that provided by the $ATP \reac
ADP + Pi$ reaction.

\subsection{Example module: Enzyme-catalysed reaction}
\label{sec:example-modul-enzyme}

\begin{figure}[htbp]
  \centering
  \SubFig{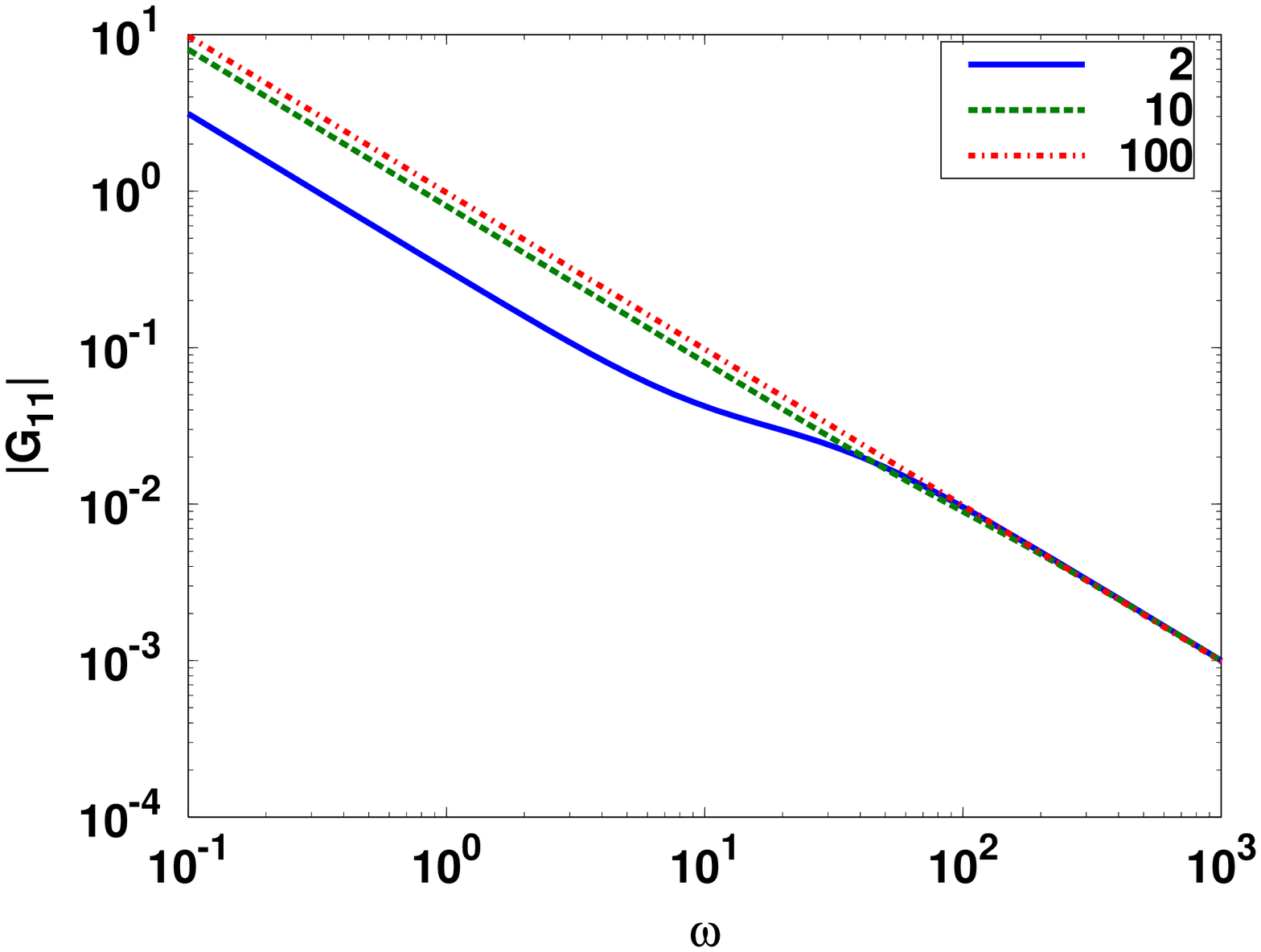}{$|G_{11}(\jw)|$}{0.3}
  \SubFig{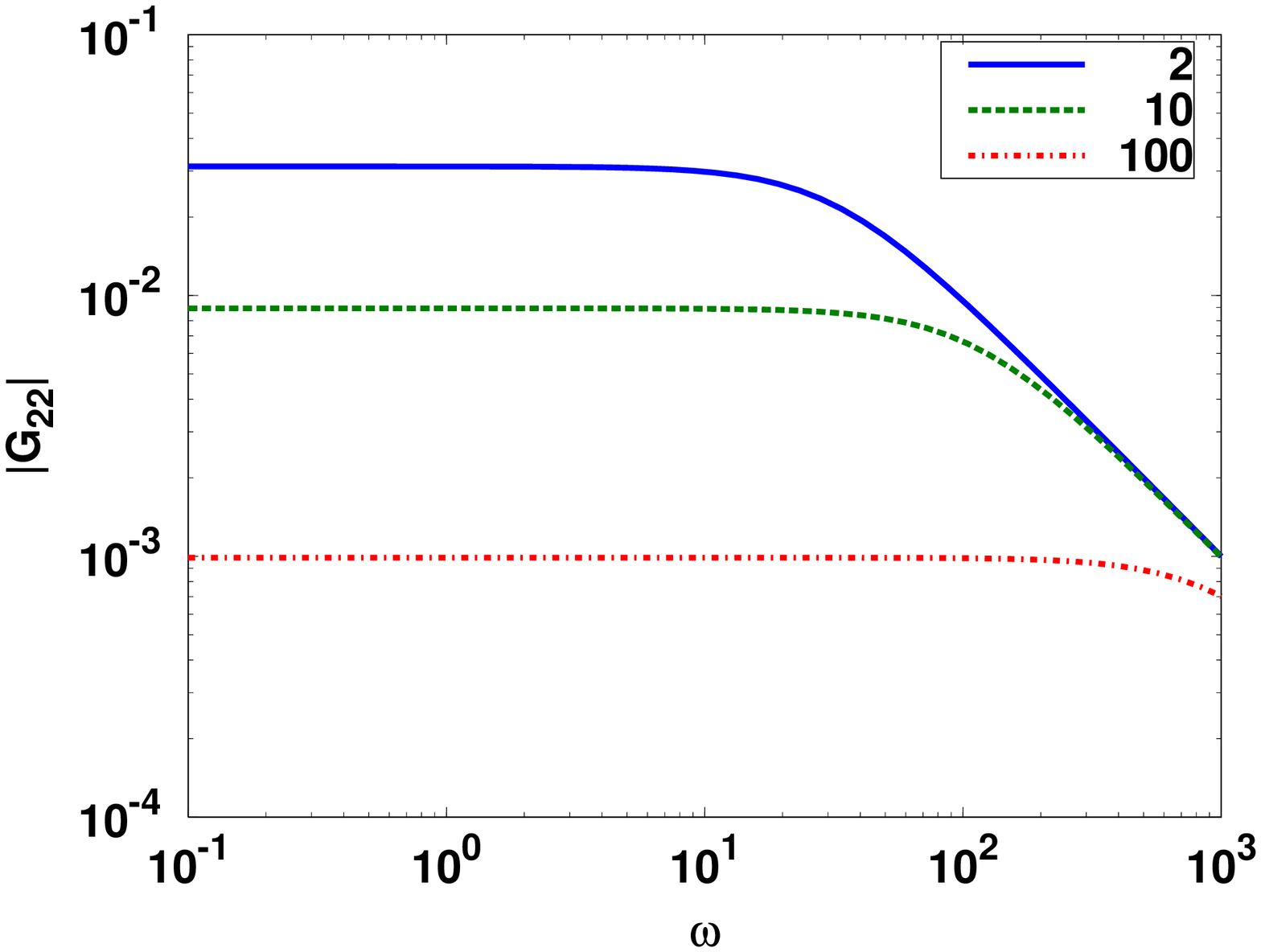}{$|G_{22}(\jw)|$}{0.3}
  \SubFig{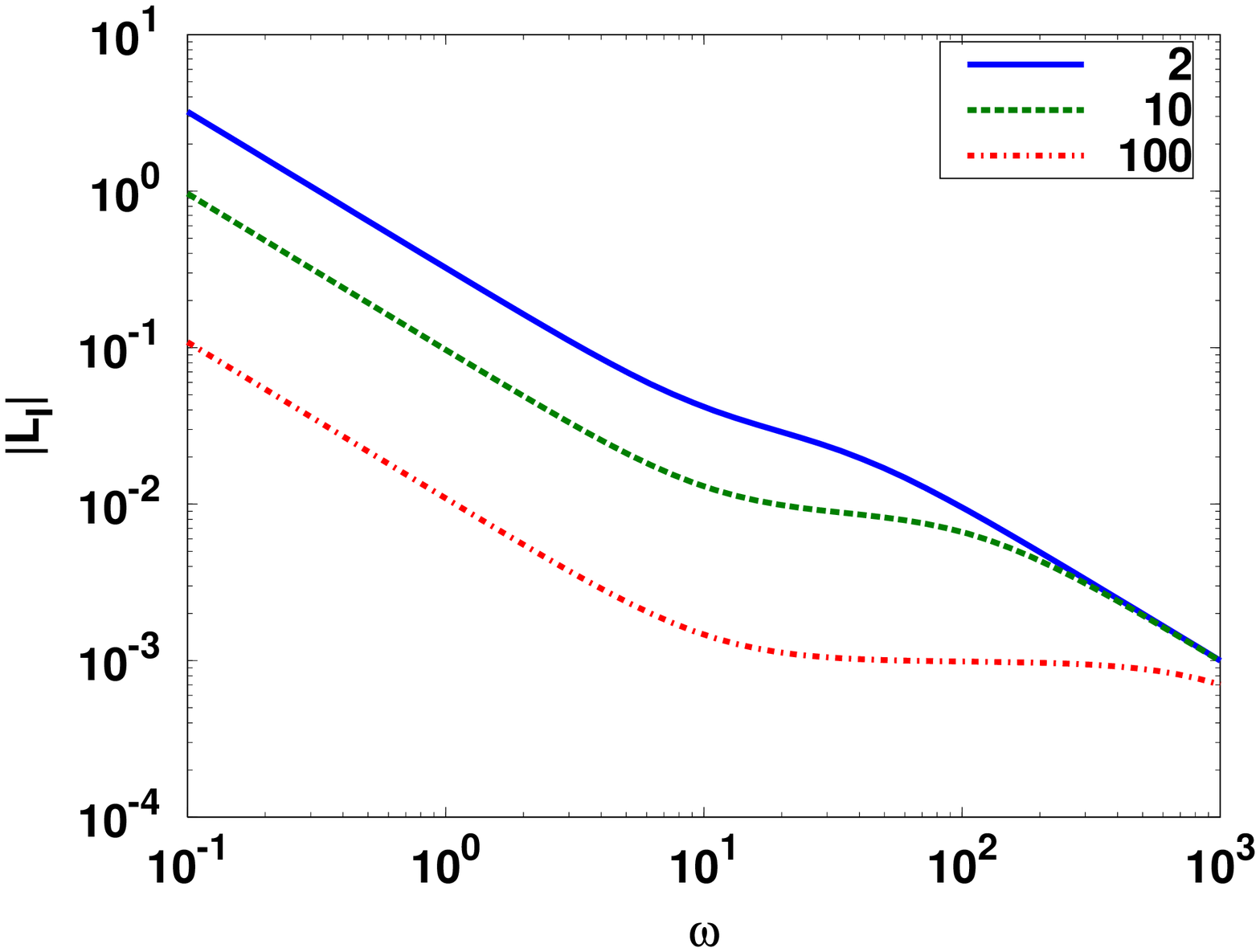}{$|L_I(\jw)|=|G_{12}(\jw)G_{21}(\jw)|$}{0.3}
  \caption{Example module: Enzyme-catalysed reaction. The three
    frequency responses are plotted for $K_A = 2,10,100$ against
    frequency $\omega\Si{~rad.sec^{-1}}$.}
  \label{fig:ECR-FR}
\end{figure}
As an example, the enzyme-catalysed reaction of Figure \ref{subfig:ABEC_abg}
is considered as a two-port module (as illustrated in Figure
\ref{fig:modularity}).  In particular, the flowstat corresponding to
\BRe{r0} is replaced by port 1 and the chemostat corresponding to
\BC{B} is replaced by port 2. Thus Equation \eqref{eq:module_tf}
becomes:
\begin{equation}\label{eq:module_tf_ECR}
  \begin{pmatrix}
    \ttt{v}_2\\ \ttt{\nmu}_E
  \end{pmatrix} =
  \begin{pmatrix}
    G_{11} & G_{12}\\
    G_{21} & G_{22}
  \end{pmatrix}
  \begin{pmatrix}
    \ttt{v}_0\\ \ttt{\nmu}_B
  \end{pmatrix}
\end{equation}
The system parameters were $K_B=K_C=K_E=1$, $\kappa_1=10$ and
$\kappa_2=1$. Three alternative values were used for $K_A$: $2$, $10$
and $100$. Using an initial state $\XX_0 = \lb100\;1\;0\;1\rb^T$, the
steady states were found for each value of $K_A$ and the
system was linearised using the method of \S~\ref{sec:linearisation}.
The transfer functions for the three cases were found to be:
\begin{align}
G_{2} &= \begin{pmatrix}
 \frac{ -s + 10}{s^2 + 32s } & \frac{ -0.34s - 10.31}{s + 32} \\
 \frac{ 2.91s + 32}{s^2 + 32s } & \frac{ -1}{s + 32} \\
\end{pmatrix}\\
G_{10} &= \begin{pmatrix}
 \frac{ -s + 90}{s^2 + 112s } & \frac{ -0.10s - 10.80}{s + 112} \\
 \frac{ 10.18s + 112}{s^2 + 112s } & \frac{ -1}{s + 112} \\
\end{pmatrix}\\
G_{100} &= \begin{pmatrix}
 \frac{ -s + 990}{s^2 + 1012s } & \frac{ -0.01s - 10.98}{s + 1012} \\
 \frac{ 92s + 1012}{s^2 + 1012s } & \frac{ -1}{s + 1012} \\
\end{pmatrix}
\end{align}
Although these transfer functions are simple enough to analyse
directly, in more complex cases it is useful to look at the transfer
function frequency responses obtained by replacing the Laplace variable
$s$ by $\jw$ where $j=\sqrt{-1}$ and $\omega$ is a frequency in
$\si{rad.sec^{-1}}$.
Figure \ref{fig:ECR-FR} gives the frequency response magnitude of the
three transfer functions: $G_{11}$ relating $\ttt{v}_0$ to $\ttt{v}_2$, $G_{22}$
relating $\ttt{\nmu}_B$ to $\ttt{\nmu}_E$ and the loop-interaction
$L_I = -G_{12}G_{21}$. for each of the three cases.

The forward transfer function $G_{11}$ approaches $\frac{1}{s}$ as
$K_A$ increases, the  transfer functions $G_{22}$ and $L_I$
decrease as $K_A$ increases. Thus larger values of $K_A$ give approximate
behavioural modularity. However, this comes at an energetic cost
measured by the external flow associated with the chemostat \BC{A}.

\subsection{Example module: Phosphorylation/dephosphorylation}
\label{sec:phosph}
\begin{figure}[htbp]
  \centering
   \SubFig{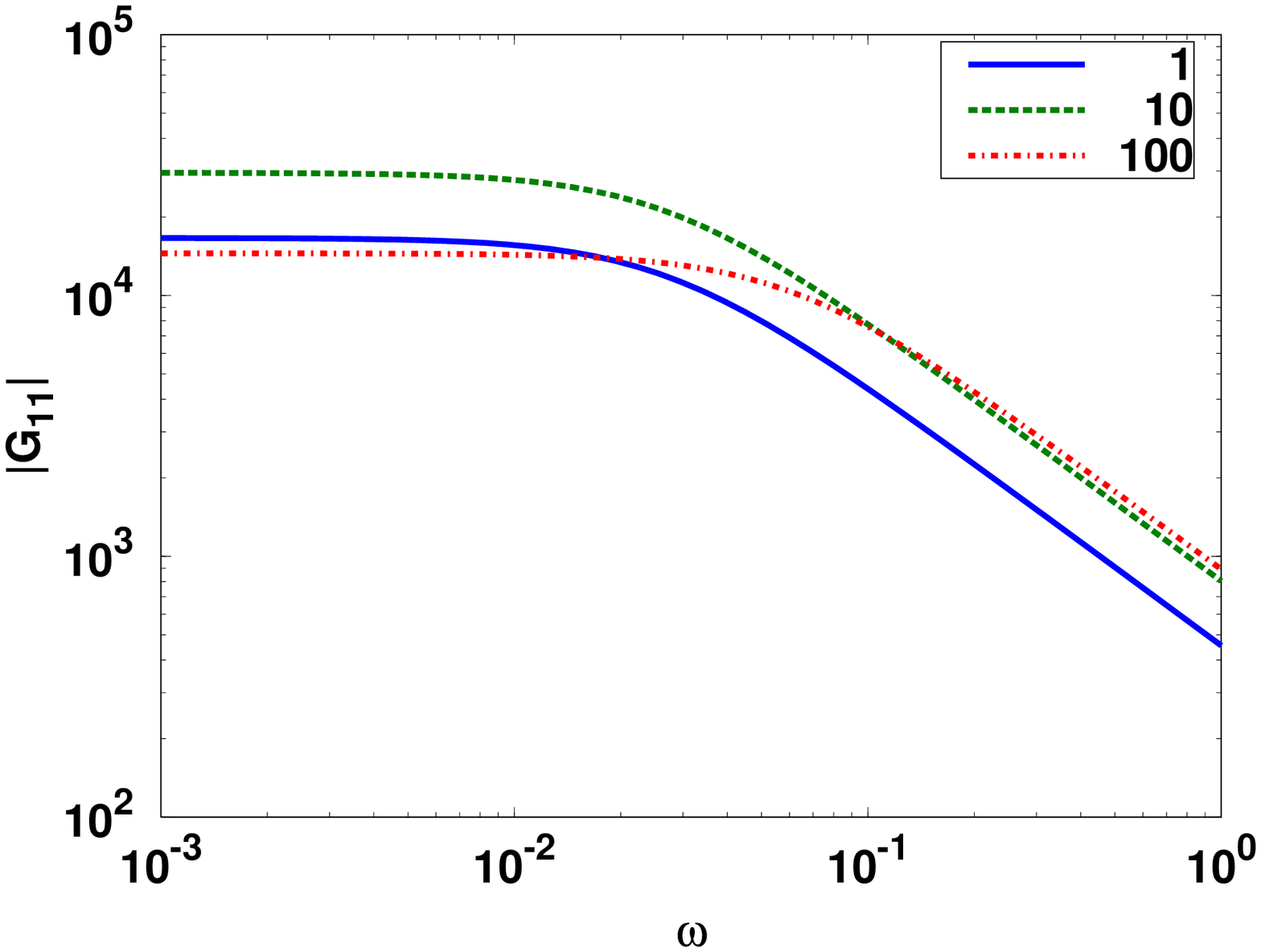}{$|G_{11}(\jw)|$}{0.3}
   \SubFig{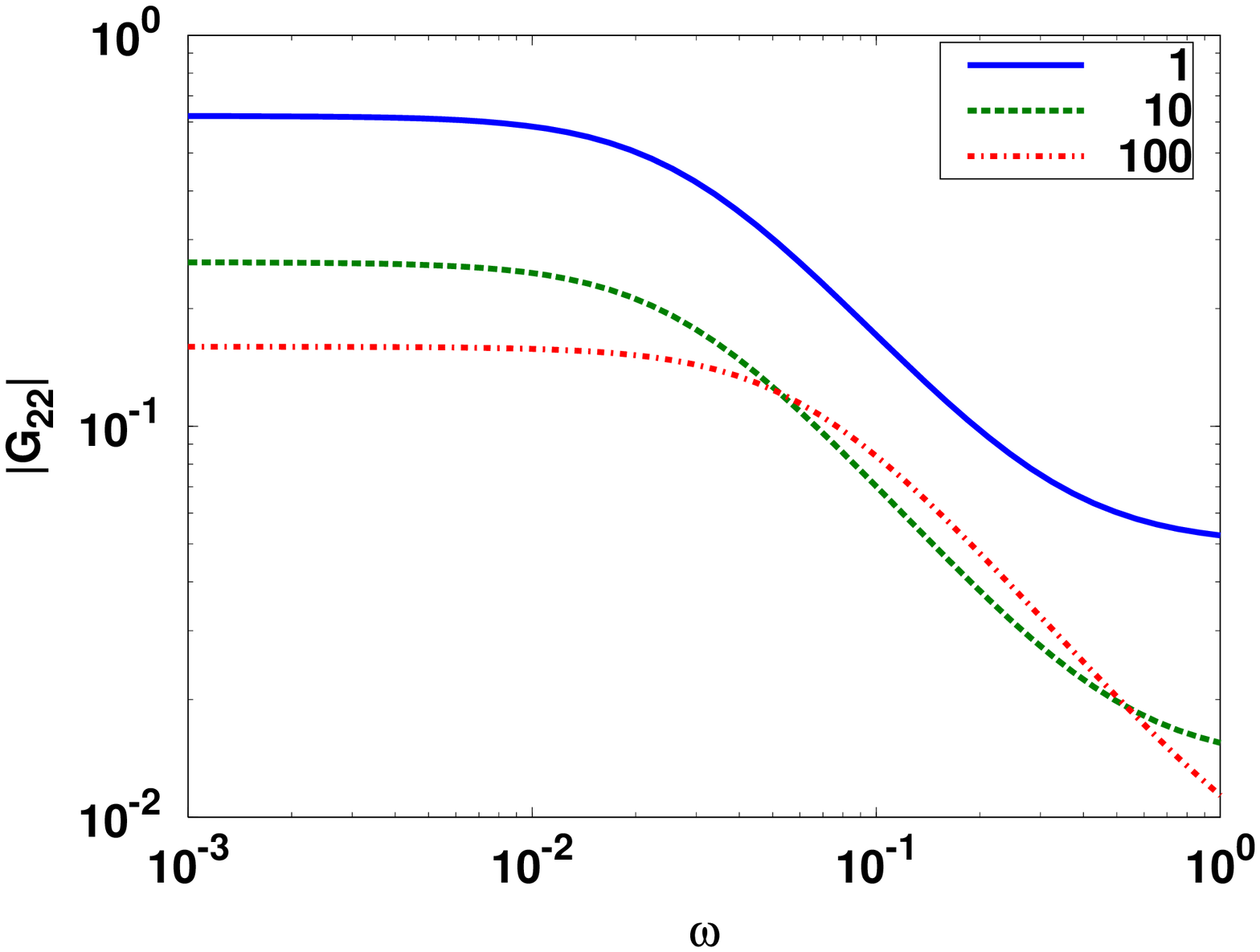}{$|G_{22}(\jw)|$}{0.3}
   \SubFig{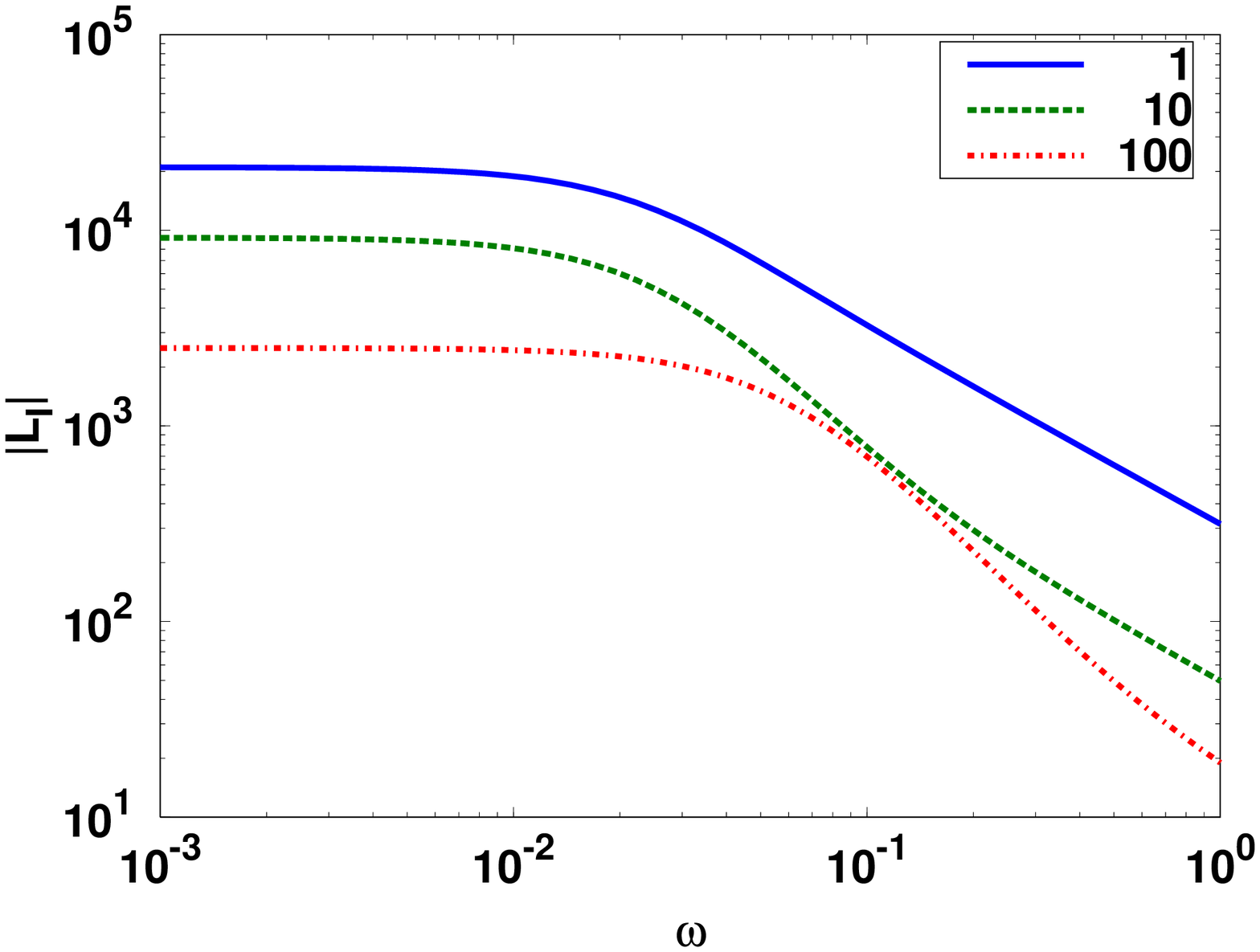}{$|L_I(\jw)|=|G_{12}(\jw)G_{21}(\jw)|$}{0.3}
  \caption{Example module: Phosphorylation/dephosphorylation}
  \label{fig:PD}
\end{figure}
A bond graph  model of the thermodynamically correct formulation of
the phosphorylation/dephosphorylation cycle of \citet{BeaQia10} was
presented by \citet{GawCra14}. Figure \ref{subfig:mPD_abg} shows a modular
version where the two ports are given by flowstat \BRe{r0} and the
chemostat \BC{MP}. The three components representing $ATP$, $ADP$ and
$Pi$ (\BC{ATP},\BC{ADP} and \BC{P}) are also chemostats and provide
the power source for the module. 

As in \S~\ref{sec:example-modul-enzyme}, this module can be analysed
by plotting the frequency response of the three transfer
functions. The parameters (which are illustrative and do not correspond
to a specific biological instance) are:
\begin{xalignat}{4}
  \XX &= \begin{pmatrix}
x_{E1}\\
 x_{C1}\\
 x_{E2}\\
 x_{C2}\\
 x_{ATP}\\
 x_{ADP}\\
 x_{P}\\
 x_{M}\\
 x_{MP}\\
 \end{pmatrix}&
\XX_0 &= \begin{pmatrix}
0\\
 0\\
 0.001\\
 0\\
 1,10,100\\
 1\\
 0.01\\
 10\\
 0\\
 \end{pmatrix}&
K &= \begin{pmatrix}
100\\
 1\\
 100\\
 1\\
 0.1\\
 0.001\\
 0.001\\
 1\\
 1\\
 \end{pmatrix}&
\kappa &= \begin{pmatrix}
10\\
 1000\\
 10\\
 1000\\
 \end{pmatrix}
\end{xalignat}
%
%
The (fixed) amount of $ATP$ was
set at three alternative values: $x_{ATP}=1,10,100$. As in
\S~\ref{sec:example-modul-enzyme}, larger values give reduced loop
interaction at the expense of more power needed to drive the module.

\subsection{Example module: Feedback inhibition}
\label{sec:FI}
The idea that a product can inhibit an enzyme and thus give negative
feedback is a well-established concept in biology
\citep{MonChaJac63,Sav09,Fel97,Cor12}. This section focuses on one
possible mechanism, competitive inhibition \citep[\S~1.4.3]{KeeSne09}.
The basic idea is that the product $P$ binds to the enzyme $E$ to form
a complex $C$ (thus partially sequestering $E$)  via the reaction:
\begin{equation}
  \label{eq:FI_chem}
  P+E \reac C
\end{equation}
Together with an additional flow of enzyme modelled by \BRe{r0}, this
reaction is modelled by the bond graph  of Figure \ref{subfig:cFI_abg}.
This can be represented as a two-port module if \BC{P}, \BRe{r0} and
associated junctions are replaced by ports. 
This module will be used in the sequel to apply feedback inhibition to
the MAPK cascade.

\section{MAPK cascades.}
\label{sec:map-kinase-cascades}
\begin{figure}[htbp]
  \centering
  \SubFig{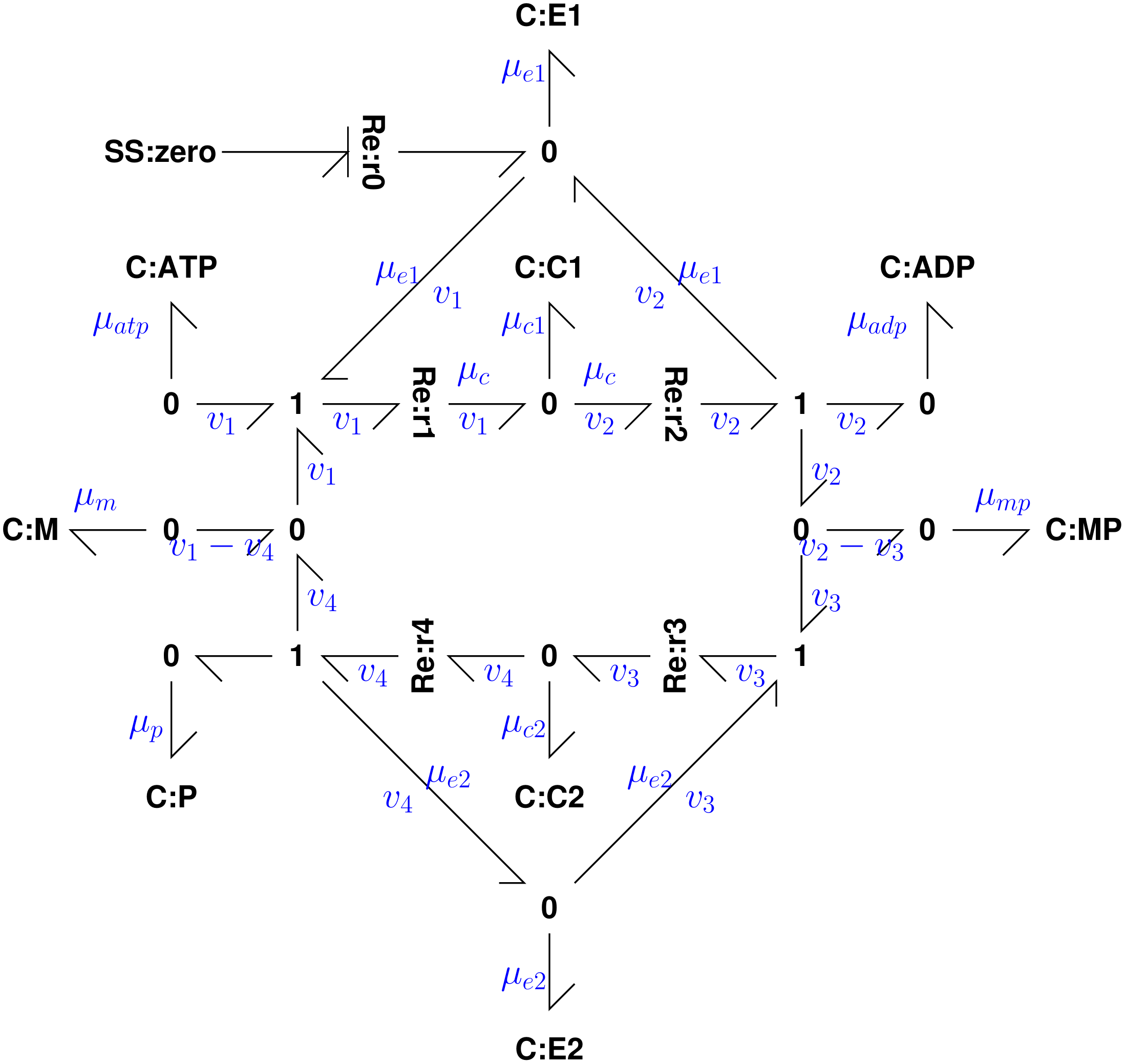}{Phosphorylation/dephosphorylation}{0.45}
  \SubFig{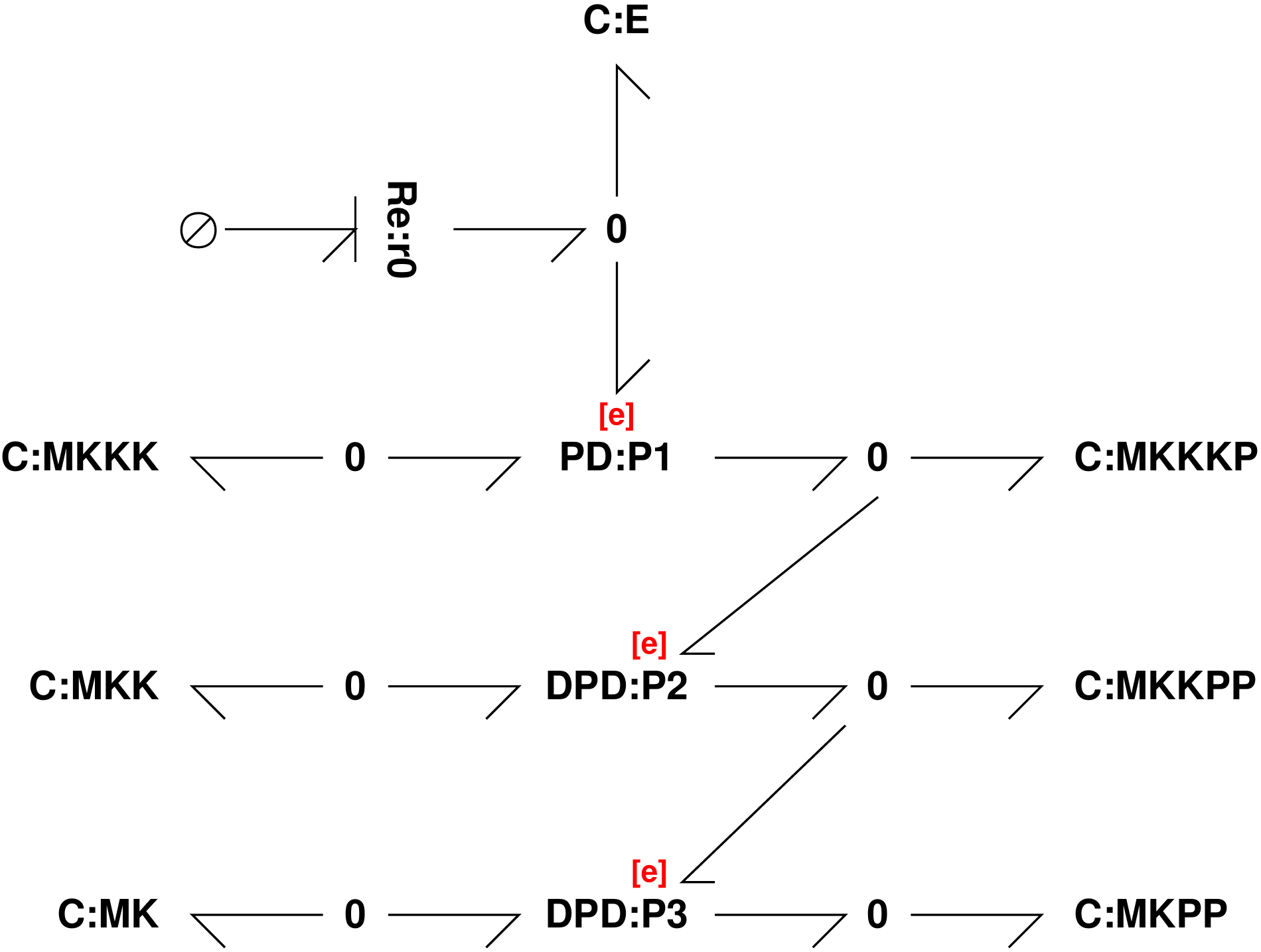}{MAPK cascade}{0.45}
  \SubFig{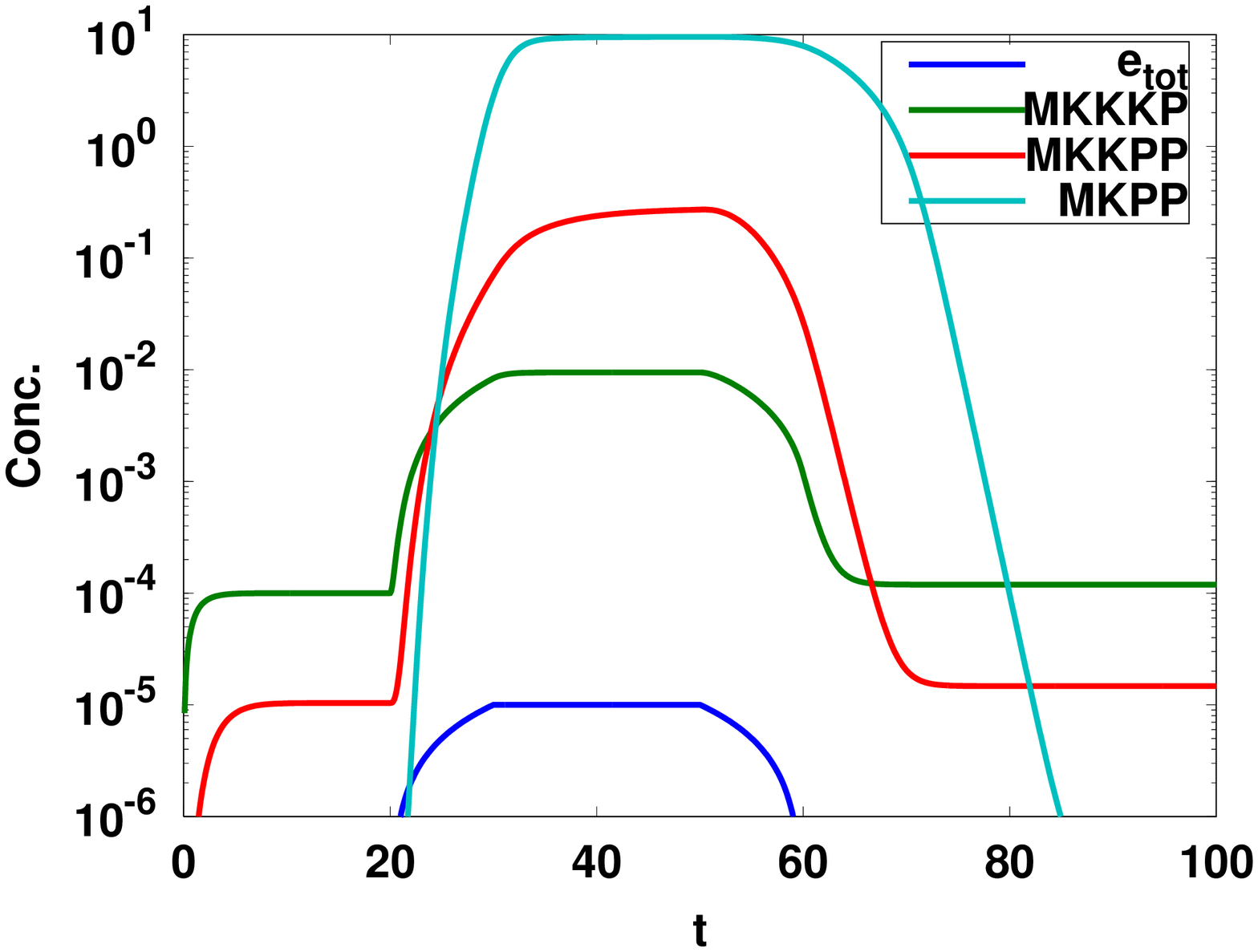}{Time courses}{0.45}
  \SubFig{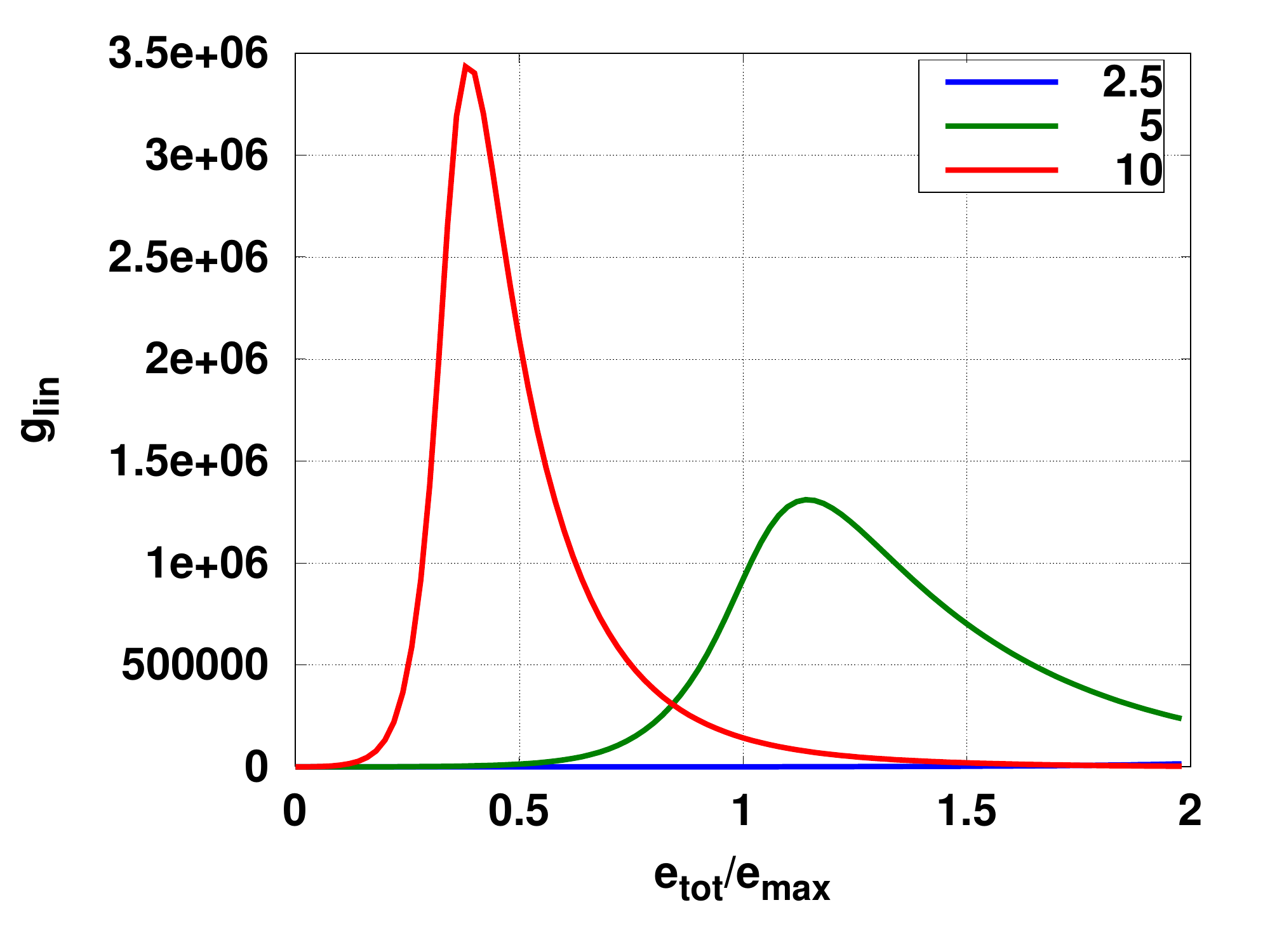}{Effect of $ATP$}{0.45}
  \caption{MAPK cascade: modular model}
  \label{fig:MAP_bg}
\end{figure}
The mitogen-activated protein kinase (MAPK) cascade is a well-studied
signalling pathway with ultrasensitive components
\citep{HuaFer96,Kho00,KolCalGil05}. However, the use of the  Michaelis-Menten
approximation to enzyme-catalysed reactions can be misleading in this
context. 
In particular, as discussed by \citet[\S 9.5]{Voi13}, ``It is tempting
to set up the two phosphorylation steps with Michaelis-Menten rate
functions, but such a strategy is not the best option, because (1) the
enzyme concentration is not constant, (2) the enzyme concentration is
not necessarily smaller than the substrate concentration, and (3) the
two reaction steps are competing for the same enzyme.''.
%

This section shows that the bond graph property of computational
modularity can be used to build a computational model of the MAPK
cascade which is thermodynamically correct and thus avoids the
pitfalls associated with inappropriate use of the Michaelis-Menten
approximation. Moreover, having seen in \S~\ref{sec:phosph} that that
the bond graph module corresponding to
phosphorylation/dephosphorylation can be designed to give approximate
behavioural modularity, the MAPK cascade can be built with approximate
behavioural modularity.

Figure \ref{subfig:MAP_abg} shows the bond graph of the MAPK cascade
based on the phosphorylation/dephosphorylation module \textbf{PD} of
Figure \ref{fig:PD} and the double phosphorylation/dephosphorylation
module \textbf{DPD} the \BRe{r0} component is
used as a flowstat generating a flow $v_0$ as discussed in Section
\ref{sec:example-modul-enzyme}. The nonlinear system of ODEs corresponding to
Figure \ref{subfig:MAP_abg}  was simulated  for 100 time units  with
an input $v_0$ given by
\begin{equation}
  \label{eq:v_0_sim}
  v_0 =
  \begin{cases}
    10^{-6} & 20\ge t \ge 30\\
    -10^{-6} & 50\ge t \ge 60\\
    0 & \text{otherwise}
  \end{cases}
\end{equation}
This gives a maximum value of the total enzyme of $e_{max}=10^{-5}$.
The system parameters are those used in \S~\ref{sec:phosph}.

Figure \ref{subfig:MAP_10_lnX} shows the corresponding time courses for
the total amount of enzyme
$e_{tot}$, and the amounts of $MKKKP$, $MKKPP$ and $MKPP$. A logarithmic scale is used to
account for the large range of values. Note that the gain between
$e_{tot}$ and the concentration $x_{MKPP}$ is of the order of $10^6$.
%

%
The steady-state value of $x_{MKPP}$ was computed for a range of
values of $e_{tot}$ and the incremental values
$\frac{dx_{MKPP}}{de_{tot}}$ were computed numerically for three
values of $ATP$: $x_{ATP} =2.5,5,10$.  Figure \ref{subfig:MAP_g_l}
shows the incremental gain plotted against $e_{tot}$.  The high gain
due to the ultrasensitivity of the phosphorylation/dephosphorylation
modules vanishes between $ATP$ amounts of $2$ and $5$.

\begin{figure}[htbp]
  \centering
  \SubFig{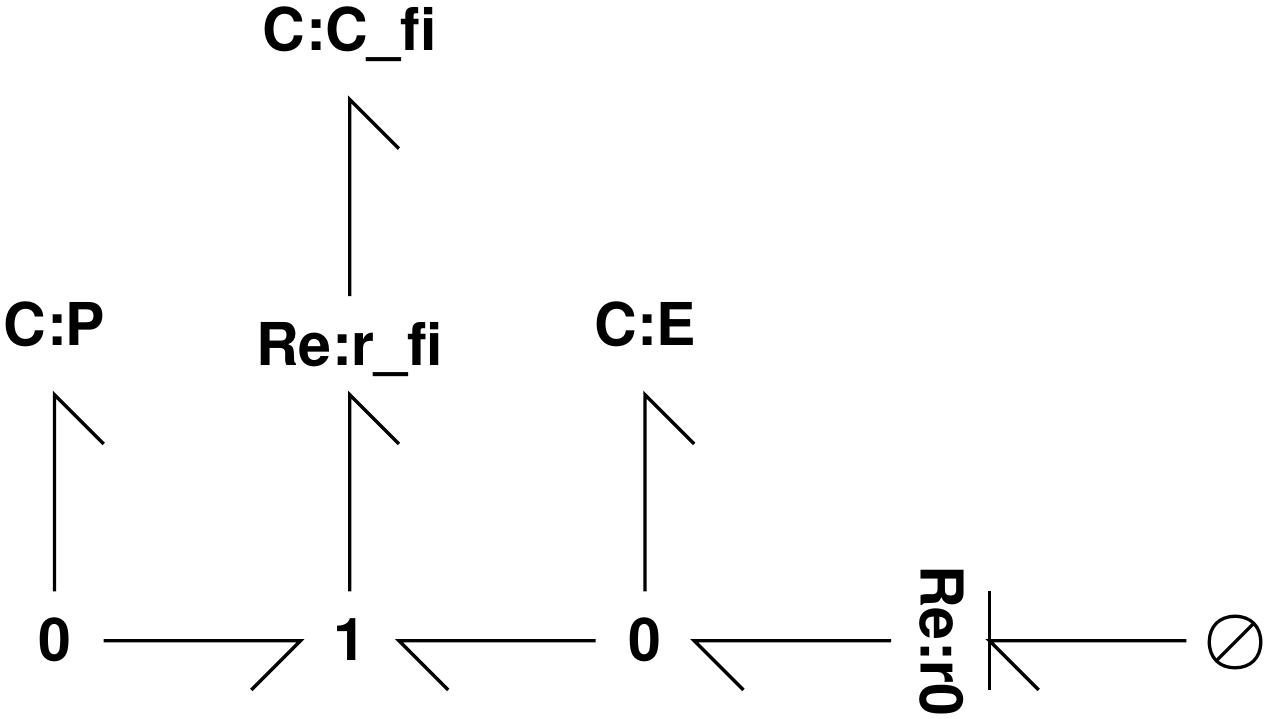}{Feedback inhibition}{0.35}
  \SubFig{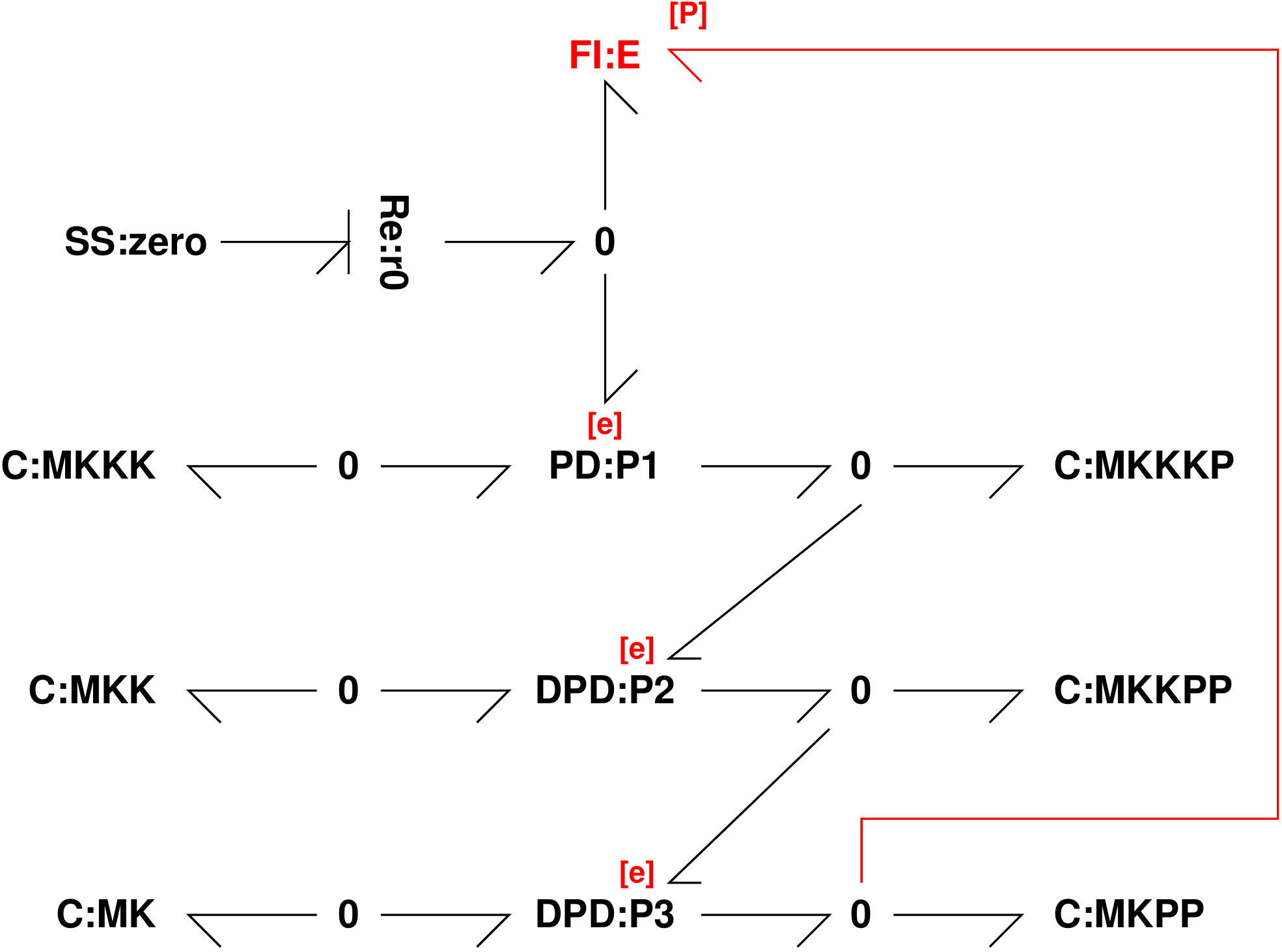}{MAPK cascade with feedback inhibition }{0.55}
  \SubFig{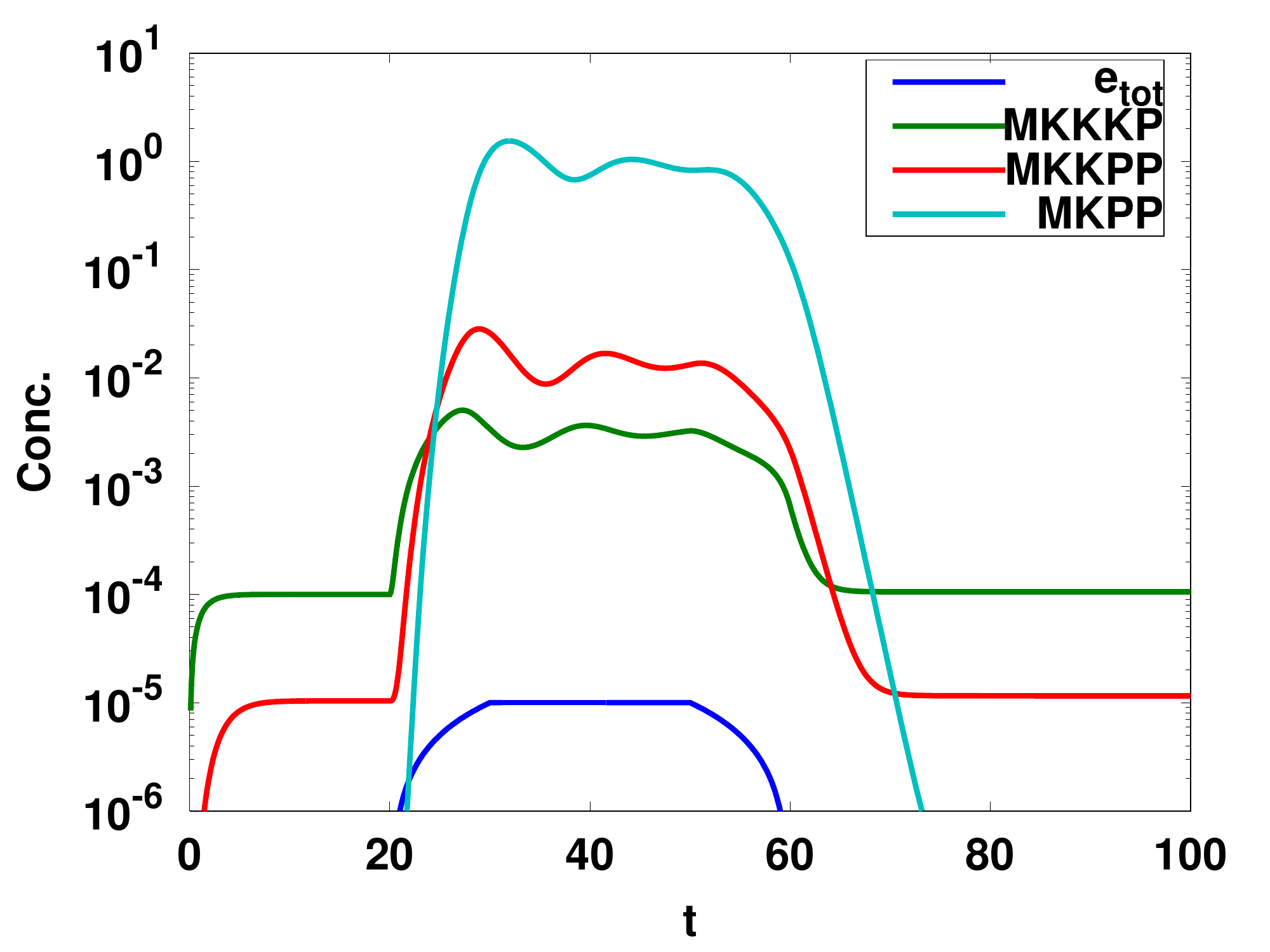}{Time courses}{0.45}
  \SubFig{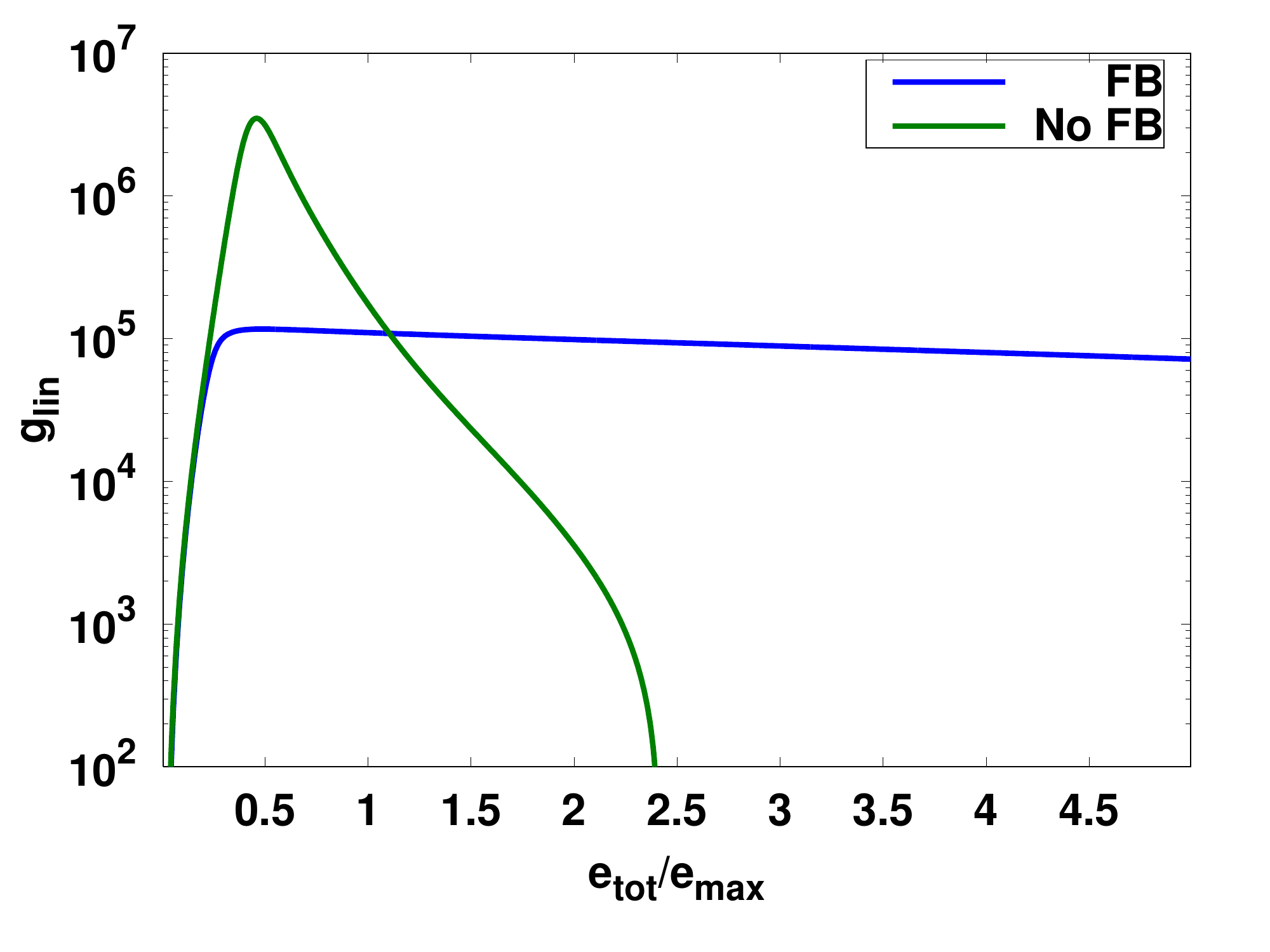}{Incremental gains.}{0.45}
  \caption{MAPK cascade with feedback inhibition}
  \label{fig:MAPFB_bg}
\end{figure}
In his seminal paper \citet{Bla34} points out that ``by building an
amplifier whose gain is deliberately made, say 40 decibels higher than
necessary ..., and then feeding the
output back on the input in such a way as to throw away the excess
gain, it has been found possible to effect extraordinary improvement
in constancy of amplification and freedom from non-linearity.''
In this context, Figure \ref{subfig:mMAPFB_abg} is the same as Figure
\ref{subfig:MAP_abg} except that the feedback inhibition module of
\S~\ref{sec:FI} is incorporated in to the bond graph and the system is
re-simulated with $K_{fi}=4$ and $\kappa_{fi}=1$. 

The steady-state value of $x_{MKPP}$ was computed for a range of
values of $e_{tot}$ and the incremental values
$\frac{dx_{MKPP}}{de_{tot}}$ were computed numerically both with and
without feedback and plotted in Figure \ref{subfig:mMAPFB_gains}.  The gain
of the system is reduced by a factor of about $20$ but the system is
now more linear: the gain is approximately constant over a wider range
of $e_{tot}$ than was the case without feedback.

\section{Conclusion}
\label{sec:conclusion}
Building on its inherent computational modularity; it has been shown
that the bond graph approach can be used to explain and adjust
behavioural modularity. The MAPK cascade was used as an example to
illustrate this point.
%
It would be interesting to repeat the MAPK examples with parameter values
taken from the literature \citep{NguMatCro13,AhmGraEdw14}. This may provide
insight into the evolutionary trade-off between energy consumption and
signalling performance \citep{HasOttCal10,AtaOrrRam04,AlbBloNew14}.

Control-theoretic concepts based on linearisation were shown to
provide a quantitative analysis of behavioural modularity.  However,
nonlinear systems can be approximated in other ways apart from
linearisation. In the context of metabolic network modelling,
\citet{Hei05} discusses and compares a number of approximations
including: logarithmic-linear, power law generalised mass action,
S-systems \citep{Sav09,Voi13} and linear logarithmic
\citep{WanKoVoi07,BerBriJon11}. It would be interesting to see whether
such approximations provide an alternative to linearisation in
analysing behavioural modularity.

It has been suggested that metabolism and its dysfunctions may related
to certain diseases including Parkinson's disease \citep{CloMidWel12,Wel12},
heart disease \citep{Neu07}, cancer \citep{MasAsg14,YizDevRog14} and
chronic fatigue \citep{MorMae13}. It is envisaged the the energy-based
approach used in this paper will help to understand such
energy-related diseases.



The example in this paper examines a signalling network as an analogy
to an electronic amplifier. Gene regulatory networks have been
analysed and synthesised as amplifiers \citep{ZhaTurYur07,BatTur07,Nak15}.
Future work will examine the bond graph based analysis and synthesis
of gene regulatory networks.

\subsection*{Acknowledgements}
Peter Gawthrop would like to thank the Melbourne School of Engineering
for its support via a Professorial Fellowship.
This research was in part conducted and funded by the Australian
Research Council Centre of Excellence in Convergent Bio-Nano Science
and Technology (project number CE140100036).

\bibliography{common}

\end{document}